\documentclass[11pt]{article}
\usepackage{graphicx}
\usepackage{amsmath}
\usepackage[latin1]{inputenc}
\usepackage[T1]{fontenc} 
\usepackage[french]{babel}
\usepackage{psfrag}

\newcommand{\be}{\begin{equation}}
\newcommand{\ee}{\end{equation}}
\newcommand{\bc}{\begin{center}}
\newcommand{\ec}{\end{center}}
\newcommand{\bea}{\begin{eqnarray}}
\newcommand{\eea}{\end{eqnarray}}
\newcommand{\ba}{\begin{aligned}}
\newcommand{\ea}{\end{aligned}}
\newcommand{\bma}{\begin{bmatrix}}
\newcommand{\ema}{\end{bmatrix}}

\newcommand{\bi}{\begin{itemize}}
\newcommand{\ei}{\end{itemize}}

\newcommand{\dila}{{\mathcal D}}

\newcommand{\tg}{\text{tg}\hspace{.02em}}

\newcommand{\degr}{^\circ}
\newcommand{\ave}[1]{\langle #1 \rangle}

\renewcommand{\author}[1]{
\begin{center}
{\bf #1}
\end{center}
\par
}
\renewcommand{\title}[1]{\begin{center}
    {\large{\bf #1}}
     \end{center}
\medskip
} 

\begin{document}
\title{  
Influence des param\`etres microm\'ecaniques dans la simulation num\'erique discr\`ete des mat\'eriaux granulaires~: assemblage, 
d\'eformation quasi-statique, \'ecoulements}
\author{
Jean-Noël ROUX  et Fran\c{c}ois CHEVOIR}
\bc
Laboratoire Navier,
Unité Mixte de Recherche LCPC-ENPC-CNRS\\
2, allée Kepler, Cité Descartes, 77420 Champs-sur-Marne, France
\ec
\bc
\begin{minipage}{16cm}
{\small
   \parindent=0pt
    {\bf AVANT-PROPOS : }
Le pr\'esent document constitue le chapitre 8 du trait\'e  {\bf Mod\'elisation num\'erique discr\`ete des mat\'eriaux granulaires}, ouvrage collectif, sous la direction
de Farhang Radja{\"\i} et Fr\'ed\'eric Dubois,  
publi\'e dans la collection << m\'ecanique et ing\'eni\'erie des mat\'eriaux >> (MIM) aux \'editions Lavoisier en 2010. Le titre, la pagination et l'indexation des r\'ef\'erences
en sont diff\'erents. 

Il dresse un bilan de l'influence 
des diff\'erents param\`etres microm\'ecaniques que l'on introduit dans la mod\'elisation des mat\'eriaux granulaires \`a l'\'echelle des grains et des contacts sur le comportement
m\'écanique macroscopique en d\'eformation quasi-statique ou en \'ecoulement. On insiste en particulier sur la d\'efinition de param\`etres de contr\^ole sans dimension, qui permettent 
une classification commode et g\'en\'erique de diff\'erents r\'egimes rh\'eologiques.

Ce texte contient
quelques r\'ef\'erences 
aux autres chapitres du m\^eme trait\'e mais peut se lire ind\'ependamment.
\par
}
\end{minipage}
\ec

\section{Introduction}
La simulation d'un mat\'eriau granulaire mod\'elis\'e \`a
l'\'echelle de ses constituants microscopiques fait intervenir de
nombreux param\`etres, li\'es \`a la description g\'eom\'etrique
du syst\`eme, aux lois d'interaction, et aux sollicitations
ext\'erieures.

Dans ce chapitre nous tentons un bilan de l'influence de ces
param\`etres sur le comportement m\'ecanique de mat\'eriaux
granulaires en nous limitant aux mod\`eles les plus simples, avec
des grains sph\'eriques ou circulaires, et en privil\'egiant les
situations qui permettent de d\'eterminer des comportements
m\'ecaniques exprim\'es par des lois constitutives. Ces lois sont
li\'ees \`a des m\'ecanismes microscopiques que nous \'evoquons
bri\`evement. Les comportements observ\'es en simulation
num\'erique peuvent se comparer quantitativement aux exp\'eriences
de laboratoire avec certains mat\'eriaux mod\`eles
 -- assemblages de billes de verre
par exemple. Avec les mat\'eriaux plus complexes, naturels ou
manufactur\'es, qui int\'eressent directement l'ing\'enieur ou le
physicien appliqu\'e, les comparaisons sont plus qualitatives.
Toutefois ces << vrais >> mat\'eriaux restent r\'egis par des lois
analogues, issues de m\'ecanismes rh\'eophysiques similaires, pour
lesquelles l'\'etude de syst\`emes mod\`eles fournit d'utiles
indications d'ordres de grandeur.

La partie~\ref{sec:param}, à partir des mod\`eles microm\'ecaniques les plus utilisés,
dresse la liste des param\`etres de contr\^ole et propose une
analyse dimensionnelle adapt\'ee aux sollicitations les plus
courantes. L'influence de ces param\`etres est ensuite discut\'ee
dans diff\'erentes conditions~: d\'eformation quasi-statique de
matériaux granulaires solides (§~\ref{sec:quasistatique}) ;
\'ecoulements granulaires denses (§~\ref{sec:ecoulement}).

\section{Simulation de mat\'eriaux mod\`eles~: param\`etres de contr\^ole
\label{sec:param}}

\subsection{G\'eom\'etrie} \label{sec:geom}
Nous traitons presque exclusivement ici d'assemblages de
particules sph\'eriques \`a trois dimensions (3D), voire
circulaires en deux dimensions (2D), avec une faible \'etendue
granulom\'etrique. Ces mod\`eles font l'objet de l'immense
majorit\'e des \'etudes num\'eriques par simulations discr\`etes.
Toutefois, la forme et la distribution de tailles des grains
jouent un r\^ole d\'eterminant dans leurs propri\'et\'es
m\'ecaniques, souvent bien davantage que certaines
caract\'eristiques microm\'ecaniques des contacts. C'est surtout
l'exp\'erience qui a r\'ev\'el\'e l'importance de ces facteurs
g\'eom\'etriques. Quelques observations issues de travaux
num\'eriques seront n\'eanmoins \'evoqu\'ees.

D'autre part, quelle que soit la forme des grains, la
g\'eom\'etrie des assemblages qu'ils forment joue un r\^ole
central dans toutes les \'etudes microm\'ecaniques. Ses
propri\'et\'es telles que la densit\'e, la connectivit\'e,
l'anisotropie \'eventuelle, la distribution des interstices entre
grains voisins sont intimement coupl\'ees au comportement
m\'ecanique du mat\'eriau, qui ne r\'esulte pratiquement jamais --
c'est l\`a une difficult\'e essentielle -- d'une moyenne simple du
comportement m\'ecanique du contact. Dans une situation
d'\'ecoulement stationnaire (§~\ref{sec:ecoulement}), la
microstructure d\'etermine la cin\'ematique du syst\`eme autant
qu'elle en est le produit. Dans un assemblage granulaire solide
(§~\ref{sec:quasistatique}), le processus de mise en place (objet
du chapitre 8) conditionne initialement la microstructure qui
influe ensuite fortement sur le comportement. Il est important,
avant d'introduire les nombreux param\`etres li\'es aux lois de
contact, de rappeler que les propri\'et\'es m\'ecaniques des
mat\'eriaux granulaires d\'ependent, autant que des propri\'et\'es
des contacts entre grains, de la mani\`ere dont ceux-ci sont
assembl\'es.

\subsection{Lois de contact\label{sec:lois}}
Les ingr\'edients des lois de contact mises en {\oe}uvre en
simulation num\'erique r\'esultent en principe d'une
mod\'elisation microm\'ecanique~\cite{Johnson85}, éventuellement
simplifiée et complétée par des hypothèses physiquement
plausibles, dans laquelle certains choix sont guidés par un souci
de faisabilité des calculs. Il convient bien s\^{u}r de s'assurer
que les propri\'et\'es auxquelles on s'int\'eresse ne sont pas
affect\'ees par un choix trop arbitraire, voire qui conduirait \`a
une violation de propri\'et\'es physiques fondamentales comme
l'objectivit\'e ou les lois de conservation. Le plus souvent, des
compromis sont adopt\'es entre la conformit\'e du mod\`ele \`a ce
qui est connu des lois de contacts et la simplicit\'e des calculs,
en exploitant les connaissances disponibles quant \`a la
sensibilit\'e des propri\'et\'es vis\'ees aux ingr\'edients
microm\'ecaniques. Dans tous les cas, il est n\'ecessaire de
s'appuyer sur des \'etudes param\'etriques (pour lesquelles
l'analyse dimensionnelle est une aide pr\'ecieuse).

Nous ne donnons ici qu'une description abr\'eg\'ee de ces lois de
contact, destin\'ee \`a faire appara\^{\i}tre la liste des
param\`etres de contr\^ole influant sur les r\'esultats des
simulations. Nous renvoyons aux r\'ef\'erences bibliographiques,
parmi lesquelles~\cite{Johnson85} est un trait\'e de la
m\'ecanique du contact et \cite{HHL98,RoCh05} contiennent des
introductions g\'en\'erales adapt\'ees \`a la simulation des
assemblages de grains, pour une pr\'esentation plus compl\`ete.

\subsubsection{Frottement}
Nous nous limitons dans un premier temps aux cas de contacts
quasi-ponctuels~: l'\'etendue de la r\'egion de contact est
suppos\'ee faible devant le diam\`etre $d$ des grains. Au point de
contact entre les grains $i$ et $j$, ceci entra\^{\i}ne que les
efforts de contact se r\'esument \`a une force ${\bf F}$, que l'on
d\'ecompose en une partie normale, $F_N$ (scalaire) et une partie
tangentielle ${\bf F}_T$ (un vecteur en 3D).

En l'absence de coh\'esion, la force normale est r\'epulsive~: on
a $F_N>0$. En g\'en\'eral on admet l'existence d'un coefficient de
frottement $\mu$ qui intervient dans la condition de Coulomb~: \be
\vert\vert {\bf F}_T \vert\vert \le \mu F_N. \label{eq:coulomb}
\ee
\subsubsection{Élasticité}
L'\'elasticit\'e du contact est bien connue dans le cas d'objets
r\'eguliers sans ar\^etes \cite{Johnson85}. La loi de Hertz relie
alors, pour des grains sph\'eriques de diam\`etre $d$ constitu\'es
d'un mat\'eriau \'elastique de module d'Young $E$ et de
coefficient de Poisson $\nu$, la force normale $F_N$  \`a la
d\'eflexion normale $h$  des surfaces en contact
(<<~interp\'en\'etration~>> apparente)~: \be
F_N=\frac{Ed^{1/2}}{3(1-\nu ^2)}h^{3/2}. \label{eq:hertz} \ee La
variation de la partie \'elastique de la r\'eaction tangentielle
${\bf F}_T$ avec le d\'eplacement relatif tangentiel
$\vec\delta$~\cite{EB96}, du fait du crit\`ere de Coulomb \'ecrit
au niveau local du vecteur-contrainte dans la petite surface de
contact~\cite{Johnson85}, ne peut en principe s'\'ecrire que de
fa\c{c}on incr\'ementale, et donne des lois fort
compliqu\'ees~\cite{THRA88}. On est souvent amen\'e \`a les
simplifier~\cite{MGJS99,iviso1,MLRJWM08}, en gardant la raideur
tangentielle $K_T(h)$, qui appara\^{\i}t dans~: \be d{\bf
F}_T=\frac{Ed^{1/2}}{(2-\nu)(1+\nu)} h^{1/2}d\vec\delta =
K_T(h)d\vec\delta, \label{eq:mindlin} \ee ind\'ependante de
$\vec\delta$  tant que la condition de Coulomb~\eqref{eq:coulomb}
reste satisfaite. Dans le cas contraire, une fois incr\'ement\'ee
${\bf F}_T $ dans le calcul, il faut la projeter sur le c\^{o}ne
de Coulomb. De plus il faut tenir compte des variations
simultan\'ees de $F_N$  et de  ${\bf F}_T$, et transporter les
forces de contact avec les mouvements d'ensemble et relatifs (y
compris roulements et pivotements) des deux objets. Pour ce faire
on ne dispose pas en fait de loi \'etablie pour tous les cas, et
on doit se contenter de r\`egles praticables qui ne violent pas
l'objectivit\'e (discut\'ee dans~\cite{KuCh06}), ni le second
principe de la thermodynamique (il ne faut pas cr\'eer de
l'\'energie \'elastique dans un cycle, ce point est \'etudi\'e
dans~\cite{EB96}). Une mise en {\oe}uvre des lois de Hertz-Mindlin
simplifi\'ees \eqref{eq:hertz}-\eqref{eq:mindlin} qui satisfait
\`a ces conditions est d\'ecrite dans~\cite{iviso1}.

Il arrive aussi souvent que l'on ne cherche pas \`a d\'ecrire
pr\'ecis\'ement l'\'elasticit\'e du contact, et que l'on choisisse
plus simplement une loi lin\'eaire unilat\'erale : \be F_N=K_N h,\
\ \ d{\bf F}_T=K_Td\vec\delta , \label{eq:linuni} \ee avec la
condition de Coulomb~\eqref{eq:coulomb}. Les raideurs normale
($K_N$) et tangentielle ($K_T$) peuvent \^etre choisies pour
respecter l'ordre de grandeur des d\'eformations \'elastiques dans
les contacts.

Dans les relations~\eqref{eq:mindlin} ou~\eqref{eq:linuni},
$\vec\delta$ représente en fait la partie élastique du déplacement
relatif tangentiel, qui ne varie que lorsque la force de contact
est intérieure au cône de Coulomb défini par~\eqref{eq:coulomb},
ou bien pénètre à l'intérieur à partir du bord~\cite{CS79,BrDi98}.

Parfois, on consid\`ere simplement les forces élastiques comme une
mani\`ere commode d'assurer l'imp\'en\'etrabilit\'e des grains et
la mobilisation progressive du frottement. Une telle attitude
implique que l'on se place implicitement dans la limite des grains
rigides, sur laquelle nous reviendrons.
\subsubsection{Forces visqueuses et coefficients de restitution\label{sec:visq}}
La partie \emph{visqueuse} de la force de contact est le plus
souvent prise lin\'eaire dans la vitesse relative~: \be
F_N^v=\alpha_N \frac{dh}{dt},\ \ \ {\bf F}_T^v=\alpha_T
\frac{d\vec\delta}{dt}, \label{eq:visco} \ee En g\'en\'eral les
coefficients d'amortissement sont assez mal connus et ils sont
rarement identifi\'es à  partir d'un mod\`ele physique. (Une
approche par la m\'ecanique du contact entre solides
visco\'elastiques a cependant \'et\'e d\'evelopp\'ee, que l'on
trouvera pr\'esent\'ee dans la r\'ef.~\cite{BSHP96} et dans bien
d'autres \'ecrits du m\^eme groupe d'auteurs. Ce mod\`ele a
\'et\'e exploit\'e dans l'\'etude des << gaz granulaires >>).

La mod\'elisation d'un choc binaire frontal avec les
relations~\eqref{eq:linuni} et \eqref{eq:visco} d\'efinit un
probl\`eme d'oscillateur harmonique amorti,
et fixe la dur\'ee $\tau_c$ de la collision entre deux grains de masse
$m$~:
\be
\tau_c=\pi\left[ \frac{2K_N}{m}-\left(
\frac{\alpha_N}{m}\right)^2\right]^{-1/2}. \label{eq:tauc}
\ee
Le
pas de temps en dynamique moléculaire est choisi comme une petite
fraction de $\sqrt{m/K_N}$. Si $\alpha_N$ dépasse la valeur
critique $\alpha_N^c = \sqrt{2mK_N}$ (pour laquelle
$\tau_c\to\infty$ dans~\eqref{eq:tauc}), le mouvement
relatif lors du choc est suramorti. Comme il y a alors deux temps
caractéristiques dans l'évolution de la déflexion normale dans le
contact, dont le rapport augmente très rapidement avec $\zeta =
\alpha_N/\alpha_N^c$, c'est une situation numériquement
défavorable, que la commodité des calculs conduit à proscrire.
Aussi, lorsque l'on souhaite, afin d'approcher rapidement des
\'etats d'\'equilibre que l'on veut \'etudier, dissiper
efficacement l'\'energie, on tend \`a choisir des coefficients
d'amortissement proches du niveau critique, mais inférieurs
($\zeta<1$). Les lois \'elastique~\eqref{eq:linuni} et
visqueuse~\eqref{eq:visco} ci-dessus se traduisent, dans une
collision isol\'ee, par un coefficient de restitution normal $e$,
fonction de $\zeta$~:
\be
e=\exp\left[\frac{\pi\zeta}{2\sqrt{1-\zeta^2}}\right].
\label{eq:restitn}
\ee
Dans le cas hertzien, on peut se
r\'ef\'erer \`a la raideur lin\'eaire tangente, $K_N(h)$, qui
varie comme $h^{1/2}$ d'apr\`es~\eqref{eq:hertz}, et appliquer les
mêmes considérations pour le choix du pas de temps (alors variable
selon le niveau de déflexion atteint) et de l'amortissement que
dans le cas linéaire. On montre que le choix d'un amortissement
\'egal, \`a chaque instant, \`a une fraction constante $\zeta$ de
la valeur critique d'un mod\`ele lin\'eaire avec raideur $K_N(h)$
donne un coefficient de restitution normal ind\'ependant de la
vitesse relative dans une collision binaire.

Le choix du coefficient d'amortissement tangentiel $\alpha_T$ est
guidé par des considérations similaires à celui de $\alpha_N$ (à
noter que sa valeur critique fait intervenir les moments d'inertie
des grains en contact). Sa relation avec un coefficient de
restitution tangentiel est un peu plus complexe en raison de la
fréquence différente des oscillations de $h$ et de $\vec\delta$
dans un choc et de la saturation possible du critère de Coulomb.

Apr\`es avoir introduit des termes visqueux dans la force de
contact, on doit se demander si on les inclut dans les composantes
$F_N$  et ${\bf F}_T$  qui doivent satisfaire l'in\'egalit\'e de
Coulomb~\eqref{eq:coulomb}. Deux attitudes sont possibles, et les
deux choix ont \'et\'e adopt\'es dans la litt\'erature. On peut
consid\'erer que l'in\'egalit\'e de Coulomb ne concerne que les
composantes \'elastiques de la force de contact. Tout se passe
alors comme si les forces visqueuses \'etaient dues \`a la
pr\'esence d'un fluide dans la r\'egion du contact (mais sans que
l'on tienne compte de la géom\'etrie pr\'ecise de son
\'ecoulement). La force normale totale peut alors \^etre
attractive lorsque les deux grains s'\'eloignent assez rapidement.
L'autre option, qui peut sembler plus logique si on attribue les
forces visqueuses à la dissipation interne au mat\'eriau qui
constitue les grains, consiste à ajouter les forces visqueuses aux
forces \'elastiques avant d'imposer
l'in\'egalit\'e~\eqref{eq:coulomb}, auquel cas il convient
d'interdire les forces attractives en limitant au besoin
l'intensit\'e de la force visqueuse.
\subsubsection{Autres mécanismes de dissipation}
La dissipation d'énergie dans les contacts peut aussi résulter
d'un comportement plastique du matériau qui constitue les grains.
Des modélisations de contacts élastoplastiques pour les objets de
forme régulière (tels que les billes métalliques) sont présentées
dans~\cite{Johnson85}, et des lois très simplifiées sont mises en
{\oe}uvre dans certaines simulations (ainsi dans la
référence~\cite{WaBr86}, ou dans \cite{LU05} où un modèle est
proposé avec plasticité et cohésion). Nous ne discuterons pas des
paramètres de ces modèles.

D'autres mécanismes de dissipation sont utilisés comme de simples
astuces nu-mériques destinées à favoriser l'approche d'états
d'équilibre, sans grand souci de vraisemblance physique (même dans
la forme des lois choisies), voire en ignorant certaines lois de
base. Ainsi, une force visqueuse est parfois mise en {\oe}uvre qui
s'oppose simplement à la vitesse de chaque particule (comme pour
un objet solide isolé baignant dans un fluide visqueux de vitesse
nulle à l'infini). Un autre mécanisme (propos\'e notamment dans certains
logiciels commerciaux) consiste à exercer sur chaque grain une
force proportionnelle à son accélération, avec un coefficient de signe opposé
à celui du produit scalaire de la force avec l'accélération.
Tout se passe donc alors
comme si la masse de chaque objet était différente dans les phases
d'accélération ou de ralentissement. Ces procédés violent la
conservation de la quantité de mouvement, et font jouer un rôle
particulier à une vitesse de référence. On ne doit donc les
utiliser qu'avec précaution, dans les cas où le bilan de quantité
de mouvement ne joue pas de rôle important. Il faut bien sûr
éviter de les mettre en {\oe}uvre pour modéliser un écoulement ou
la mise en place d'un matériau granulaire sous gravité.
\subsubsection{Coh\'esion}
C'est l'énergie  par unité d'aire $\gamma$ des interfaces entre
les grains solides et le milieu ambiant qui est à l'origine de
l'adhésion aux contacts, qu'elle soit ou non transmise par un
ménisque liquide. La force attractive maximale est toujours
d'ordre $\gamma d$. En présence d'un pont liquide, il est possible
d'utiliser des modèles quantitatifs pour les assemblages de
billes~\cite{CMP02,RiEYRa06}. La portée $D_0$ des forces
attractives à distance est alors déterminée par la distance de
rupture, de l'ordre du volume du ménisque élevé à la puissance
$1/3$. La prise en compte de l'adhésion capillaire dans la
simulation des matériaux granulaires fait l'objet du chapitre~12.
Dans le cas de grains solides secs, la force adhésive est sensible
aux irrégularités de leur surface, et sa portée (celle des forces
de van der Waals) est de quelques nanomètres. C'est pourquoi elle
se manifeste seulement en pratique pour les particules de petite
taille, typiquement de l'ordre du micron ou de la dizaine de
microns. Pour des grains de sable, ou des billes dont le diamètre
est de l'ordre de 100~$\mu$m, seules quelques aspérités peuvent
être sensibles à une telle attraction, dont l'effet se trouve
ainsi fortement réduit, alors que les forces capillaires en
présence d'un liquide ne sont pas sensibles à ces irrégularités
géométriques et entrent en jeu pour des  grains plus gros. Il
existe sur la modélisation des effet simultanés de l'adhésion et
de la déformation élastique au contact une littérature
fournie~\cite{MAU00}, mais l'étude des assemblages granulaires
cohésifs par simulation numérique (exposée au chapitre 11) en a
pour l'essentiel retenu des versions assez simplifiées, car les
approches classiques restent incomplètes (on ne sait pas toujours
bien prendre en compte les efforts tangentiels), s'appliquent
difficilement à des grains à surface irrégulière et font appel à
des informations détaillées sur les matériaux et les états de
surface qui ne sont pas toujours disponibles en pratique. Les lois
simples mises en {\oe}uvre dans les exemples simulations rapportés
aux §~\ref{sec:quasistatique} et~\ref{sec:ecoulement} consistent à
ajouter un terme attractif simple~\cite{Rognon06a,Gilabert07}, à
la répulsion élastique exprimée par~\eqref{eq:linuni},
introduisant ainsi les deux paramètres importants communs à tous
les modèles de cohésion~: une traction maximale $F_0$ que peut
supporter un contact, et la portée $D_0$ de l'attraction mutuelle
entre surfaces solides. \`A l'équilibre, en l'absence d'action
extérieure, une paire isolée de grains est maintenue en contact
avec une déflexion \be h_0 \sim F_0/K_N, \label{eq:defh0} \ee pour
laquelle la force de répulsion élastique et la force d'adhésion se
compensent.

En présence d'adhésion, il est important d'appliquer l'inégalité
de Coulomb~\eqref{eq:coulomb} en excluant de $F_N$ au second
membre le terme attractif. Ainsi, dans un modèle simple où ce
terme est constant, égal à $-F_0$ pour les grains en
contact~\cite{Gilabert07}, on impose la condition $\vert\vert {\bf
F}_T\vert\vert \le \mu (F_N+F_0)$. En effet, il faut considérer
que les surfaces sont pressées l'une contre l'autre par
l'attraction des deux grains, qui agit de façon analogue à une
force extérieure. Il en résulte qu'en l'absence de force
extérieure, le contact entre deux grains collés avec la force
$F_0$ est alors capable de transmettre une force tangentielle
maximale égale à $\mu F_0$, alors que la force normale \emph{totale}
est nulle.

\subsubsection{Résistance au roulement}
Si on prend en compte l'extension finie des régions de contact,
qui ne se réduisent pas à des points, soit en raison de la
déformablité des grains solides, soit de leur forme irrégulière et
non convexe, on est amené à autoriser les contacts à transmettre
un moment $\Gamma$, qui travaille dans les mouvements relatifs de
roulements sans glissement (différence de taux de rotation autour
d'un axe tangentiel) ou de pivotement sans glissement (différence
de taux de rotation autour de l'axe normal). Nous évoquerons
quelques exemples de l'effet d'un tel ingrédient micromécanique,
encore relativement peu usité, dans des exemples bidimensionnels
(pas de pivotement). Pour prendre en compte ce phénomène les
modèles retenus en simulation sont restés fondés sur des approches
simplificatrices~\cite{IWOD98}. Le paramètre essentiel est le
coefficient de frottement de roulement, $\mu_R$, tel que le moment
maximal soit $\mu_R F_N$. Ainsi défini, $\mu_R$ a la dimension
d'une longueur, et c'est pourquoi on rencontre aussi la définition
dans laquelle c'est $\mu_R$ divisé par un rayon de grain $R$ qui
est appelé coefficient de frottement. La réponse d'un contact
sollicité en roulement, avant que $\Gamma$ atteigne sa valeur
limite, peut être modélisée comme purement élastique, et on
obtient alors un modèle analogue à celui de la loi de contact
tangentielle~\cite{TOST02,Gilabert07} élastoplastique habituelle.
Dans le cadre d'un modèle de grains rigides, on peut aussi étendre
l'approche de la dynamique des contacts au cas de la résistance au
roulement~\cite{Estrada08}.

Si le diamètre de la région du contact est $l$ alors $\mu_R$ doit
être de l'ordre du diamètre $l$ de la région de contact
(distance entre aspérités), tandis que la raideur en rotation relative doit
être d'ordre $K_Nl^2$. En présence de cohésion, si on écrit,
comme pour la condition de Coulomb ordinaire, l'inégalité
$\Gamma\le \mu_R F_N^r$ avec la seule composante répulsive
élastique de la force normale, on obtient des contacts capables de
transmettre un moment d'ordre $\mu_R F_0$ même si la force normale
totale est nulle. Un faible coefficient de frottement de roulement
peut donc avoir une influence importante dans le cas
cohésif~\cite{Gilabert07}.
\subsubsection{Cas des grains rigides}
La liste des paramètres que nous avons introduits jusqu'ici se
trouve plus réduite dans le cas d'une modélisation (comme en
dynamique des contacts, \emph{cf.} le chapitre 3) avec des grains parfaitement rigides.
Ainsi toutes les raideurs et les échelles de longueur associées à
une déflexion élastique disparaissent. Restent le coefficient de
frottement, éventuellement le frottement de roulement, et les
coefficients de restitution, ainsi que les données ayant le sens
d'une force, comme la traction maximale $F_0$, et éventuellement
la portée $D_0$ des forces attractives.

Toutefois, comme la condition de stricte impénétrabilité n'est
forcément satisfaite qu'approximativement dans la pratique, la
méthode introduit des niveaux de tolérance. Elle fait également
appel au choix d'un pas de temps. Nous renvoyons au chapitre 3
pour le rôle de ces paramètres liés à la mise en {\oe}uvre
numérique de la dynamique des contacts.

\subsection{Analyse dimensionnelle\label{sec:anadim}}
L'analyse dimensionnelle, fondée sur l'indépendance des propriétés
physiques vis-à-vis du choix des unités fondamentales de temps, de
longueur et de masse, nous assure que les propriétés du matériau,
si elles sont exprimées par des grandeurs sans dimension (rapport
de contraintes fonction d'une déformation par exemple), ne
dépendent que des combinaisons sans dimension des paramètres du
problème.

Dans un essai mécanique, les paramètres, en plus de ceux du
matériau, comprendront le plus souvent une pression $P$
(contrainte de confinement) et un certain taux de déformation
$\dot\epsilon$.

Bien entendu, on peut construire de bien des façons différentes
une liste de nombres sans dimension à partir des données. Le choix
que nous proposons est guidé par les considérations suivantes. Ces
nombres, d'abord, expriment, comme il est accoutumé, le rapport de
deux grandeurs physiques, par exemple à partir des échelles de
temps : le temps de collision $\tau_c$ (qui est comparable au
temps de propagation de la quantité de mouvement à travers un
grain), le temps de cisaillement $\tau_s = 1/\dot \epsilon$ et le
temps inertiel $\tau_i = \sqrt{m/P}$ (le temps de déplacement
caractéristique d'un grain de masse $m$ soumis à la pression $P$).
En présence de modèles de matériaux différents, on peut donner des
définitions assez similaires des paramètres de contrôle
adimensionnés qui permettent de situer dans l'espace des
paramètres des régions où les comportements seront comparables. Il
convient ensuite, afin de pouvoir faire appel le plus souvent
possible aux mêmes définitions pour les modèles de grains rigides
ou déformables, d'isoler les paramètres de raideur en les faisant
intervenir dans le moins de combinaisons possibles. Enfin, il est
préférable que les propriétés du système étudié possèdent des
limites bien définies pour les valeurs extrêmes, très faibles ou
très élevées, des nombres choisis.
\subsubsection{Grandeurs liées au matériau}
Le comportement des contacts fait directement intervenir des
grandeurs sans dimension~: le coefficient de frottement $\mu$, $e$
(ou $\zeta$) pour la dissipation visqueuse ou la restitution
normale, $e_T$ (ou $\zeta_T$), son homologue pour la force
tangentielle, et éventuellement le niveau de frottement de
roulement $\mu_R/d$.
\subsubsection{Raideurs adimensionnées}
Les raideurs normale et tangentielle se comparent entre elles, ce
qui donne le paramètre $K_T/K_N$ pour l'élasticité
linéaire~\eqref{eq:linuni}. Pour le modèle simplifié
\eqref{eq:hertz}-\eqref{eq:mindlin} de l'élasticité de
Hertz-Mindlin, ce rapport de raideurs est indépendant du la force
et vaut $\dfrac{2-2\nu}{2-\nu}$, qui varie entre 2/3 et 1 pour les
matériaux ordinaires dont le coefficient de Poisson $\nu$ varie
entre 1/2 et zéro. Il s'agit donc d'un paramètre dont les
variations sont très restreintes quand on cherche à reproduire les
propriétés des matériaux habituels. La comparaison de la raideur
normale $K_N$ avec le niveau de contrainte $P$, permet de définir
un \emph{niveau de raideur} $\kappa$, tel que l'on ait pour la
déflexion moyenne $h$ dans les contacts~: \be \frac{h}{d} \propto
\kappa^{-1}. \label{eq:kappa} \ee La force normale moyenne
$\ave{F_N}$ étant proportionnelle à $Pd^{D-1}$ en dimension $D$
égale à 2 ou 3, on prend \be \ba
\kappa & = \frac{K_N}{Pd^{D-2}}\  &\mbox{(cas linéaire, D=2 ou 3)}\\
\kappa &= \left(\frac{E}{P(1-\nu^2)}\right)^{2/3} \ \ &\mbox{(cas
hertzien, D=3)} \ea \label{eq:defkappa} \ee En fait, on sait
relier le rapport $\ave{F_N}/Pd^{D-1}$ à la densité et au nombre
de coordination~\cite{RiEYRa06,Gilabert07,iviso1} et on montre
alors que dans les conditions habituelles (faible étendue
granulométrique) on a un coefficient de proportionnalité proche de
1 dans~\eqref{eq:kappa}.

La limite des grains rigides est celle où $\kappa\to+\infty$. Dans
cette limite, si elle est bien définie, les propriétés du matériau
ne doivent plus dépendre de $\kappa$, et les petites déflexions
aux contacts, d'ordre $\kappa^{-1}$, ne modifient pas sensiblement
la structure de l'assemblage granulaire par rapport à celle d'un
assemblage de grains indéformables.

\subsubsection{Nombre d'inertie}

Le rapport des échelles de temps imposées par la sollicitation et
associées à la dynamique des grains de masse $m$ soumis à des
forces d'ordre $Pd^{D-1}$ définit le \emph{nombre d'inertie}, \be
I = \dot\epsilon \sqrt{\frac{m}{Pd^{D-2}}}. \label{eq:defI} \ee
(Une définition très proche, dont l'écriture est identique à 2 ou
3 dimensions, consiste à prendre $I = \dot\epsilon d
\sqrt{\frac{\rho_m}{P}}$, où $\rho_m$ est la masse volumique du
matériau constituant les grains). La limite de l'évolution
quasi-statique, dans laquelle on s'attend à ce que le système
reste proche d'un état d'équilibre à chaque instant, est la limite
où $I\to 0$.
\subsubsection{Paramètres adimensionnés liés à la cohésion}
L'effet principal des forces attractives est d'introduire une
échelle de force, $F_0$, qui se compare aux forces liées à la
contrainte de confinement. On définit ainsi un nombre de cohésion
$\eta$ qui caractérise l'intensité des forces adhésives relativement aux
forces de confinement, ou bien une pression réduite $P^*$~:
\be
\eta = \frac{F_0}{d^{D-1}P}, \ \ \ \ \  P^* = \frac{1}{\eta}=\frac{d^{D-1}P}{F_0}.
\label{eq:defeta}
\ee
Lorsque $P^*\gg 1$, les forces dues au
confinement dominent et les effets de la cohésion sont
négligeables. La structure de l'assemblage est similaire à celle
d'un système sans cohésion. Dans le cas contraire, $P^*\ll 1$, la
cohésion peut stabiliser des structures très différentes, en
particulier beaucoup plus lâches. De plus, c'est alors la
cohésion qui détermine le niveau typique des déflexions dans les
contacts, qui est d'ordre $h_0$ définie en~\eqref{eq:defh0}, et
des forces normales, alors d'ordre $F_0$ (positives ou négatives).
Lorsque la cohésion domine, plutôt que d'évaluer l'effet
de l'inertie avec le nombre $I$~\eqref{eq:defI}, on peut définir
\be
I_a = \dot\epsilon \sqrt{\frac{md}{F_0}},
\label{eq:defIa}
\ee
où la force typique $Pd^{D-1}$ est remplacée par $F_0$.

L'autre nombre sans dimension introduit par les modèles de
cohésion est le rapport $D_0/d$, pour lequel on ne rapportera pas
ici de résultat d'étude paramétrique.
\subsubsection{Bilan et remarques.}
Il faut donc étudier, plutôt que l'influence des 8 grandeurs
dimensionnées $m$, $d$, $K_N$ (ou $E$), $K_T$, $\dot\epsilon$,
$P$, $F_0$, $D_0$ sur le comportement mécanique, celui des 5
paramètres de contrôle sans dimension $\kappa$, $I$, $P^*$ (ou
$\eta$), $K_T/K_N$, $D_0/d$, auxquels il faut adjoindre les
coefficients de frottement et de restitution. Si l'influence des
rapports $K_T/K_N$ et $D_0/d$ se révèle assez secondaire, les
trois nombres $\kappa$, $I$, et $P^*$ jouent un rôle central,
ainsi que les coefficients de frottement, $\mu$ et $\mu_R/d$,
comme il est montré dans les résultats rappelés dans les
§~\ref{sec:quasistatique} et \ref{sec:ecoulement} ci-dessous. Si
on s'intéresse au comportement mécanique macroscopique, on doit
provilégier l'étude des systèmes homogènes dans la limite où le
nombre $N$ de grains tend vers l'infini, et $N$ peut être
considéré comme un autre paramètre à faire varier. Comme la durée
des calculs est proportionnel à $N$ et au nombre de pas de temps
en dynamique moléculaire, le coût de la simulation d'un intervalle
de déformation donné lorsque le taux $\dot\epsilon$ est imposé
sous un niveau de contrainte $P$ varie comme
$\dfrac{N}{I}\sqrt{\kappa}$. D'où l'importance pratique, si on
s'intéresse à la limite quasi-statique, et à la limite des grains
rigides, de savoir quelle valeur de $I$ sera assez faible et
quelle valeur de $\kappa$ sera assez élevée.

\section{Déformation de matériaux granulaires solides\label{sec:quasistatique}}

\subsection{Assemblage, confinement, compression \label{sec:comp}}
Les matériaux granulaires à l'état solide sont le plus souvent
étudiés sous un certain niveau de confinement. Celui-ci varie
typiquement de quelques kPa à quelques MPa dans les applications à
la mécanique des sols, ce qui correspond au poids d'une couche de
sable dont l'épaisseur va de quelques centimètres à plusieurs
centaines de mètres. Pour des billes de verre, en prenant
$E=70$~GPa et $\nu=0.3$ dans~\eqref{eq:hertz}, $\kappa$,
d'après~\eqref{eq:defkappa}, décroît de
181000 pour $P=1$~kPa à 1810 pour $P=1$~MPa.

L'état d'un matériau sous une pression de confinement $P$ dépend
en général à la fois du processus d'assemblage et de l'histoire de
sa compression. Les méthodes numériques d'assemblage, dans
lesquelles un matériau solide est constitué, soit par condensation
d'un gaz granulaire, soit par un algorithme géométrique, font
l'objet des chapitres 7 et 8 du présent ouvrage. Les méthodes
mécaniques, dans lesquelles on simule effectivement la trajectoire
des grains pendant l'assemblage, font apparaître d'autres nombres
sans dimension, et produisent des résultats qui peuvent dépendre
des paramètres qui auront par la suite peu d'influence (ou aucune
influence) sur le comportement quasi-statique. Ainsi la densité
des échantillons obtenus par pluviation contrôlée~\cite{Emam06}
est sensible aux coefficients de restitution normal et tangentiel,
et la connectivité des structures lâches obtenues par un processus
d'agrégation balistique isotrope de grains cohésifs dépend de la
vitesse moyenne quadratique initiale des particules, qu'il faut
comparer à la vitesse d'éloignement minimal pour échapper à
l'attraction d'une particule voisine, qui est d'ordre
$\sqrt{D_0F_0/m}$~\cite{Gilabert07}. L'assemblage par compression
homogène à partir d'une configuration lâche et sans contact peut
être obtenu numériquement, en particulier  en recourant aux
conditions aux limites périodiques décrites au chapitre 6. S'il
est assez graduel, c'est-à-dire si la valeur maximale de $I$ reste
assez faible (de l'ordre de $10^{-4}$), alors, avec des grains non
cohésifs et un niveau de raideur assez élevé ($\kappa > 5000$) ce
procédé conduit à des configurations d'équilibre sous pression
contrôlée qui ne dépendent que du coefficient de frottement $\mu$
(si $\mu_R=0$)~\cite{iviso1}.

La compression quasi-statique isotrope de matériaux non
cohésifs (c'est-à-dire la poursuite du processus au-delà de la formation
d'un assemblage solide)  entraîne une évolution modérée et
progressive~\cite{iviso2} de la microstructure du système.
Il en est de même
pour tout trajet de chargement tel que le rapport des contraintes reste
constant, comme la compression {\oe}dométrique.
Dans un assemblage de billes identiques, si le nombre de coordination
$z$ est élevé, c'est-à-dire proche de 6,  sous faible pression ($\kappa\to\infty$), il varie
d'environ 5\% quand $\kappa$ reste supérieur à 2000~; s'il est
initialement faible, proche de 4 (le plus souvent, plus près de
4,5 si on élimine les grains <<~flottants~>> \cite{iviso1}, qui
représentent alors souvent plus de 10\% du total), il augmente
plus vite, approchant 4,8 pour $\kappa=2000$ (évolution facilitée
par le recrutement d'une bonne moitié des  <<~flottants~>>
initiaux par la structure qui porte les forces). Pour $\kappa$ de
l'ordre de 1000, on constate une augmentation considérablement
plus rapide du nombre de coordination avec la pression. On
s'attend donc à un comportement proche de la limite rigide pour
$\kappa > 2000$ si $z$ est élevé, et pour des $\kappa$ un peu plus
grands si $z$ est faible. C'est la création de nouveaux contacts là où
il y a avait sous pression plus faible de petits interstices entre
grains voisins qui augmente la coordinence.
On doit donc conclure que
{\em pour approcher de la limite des grains rigides il ne suffit pas
de satisfaire à la condition $\kappa\ll 1$ que la déflexion
élastique $h$ des contacts soit faible devant le diamètre $d$ des
grains. Il faut en plus que ces déflexions n'entraînent
que de faibles modifications de la distribution des interstices
entre grains voisins.}
D'où la nécessité de satisfaire une inégalité plus exigeante, qui demande un niveau de raideur $\kappa$
minimal, dans la pratique des systèmes simples considérés dans ce chapitre, de l'ordre de quelques milliers.
\begin{figure}[!htb]
\centering
\includegraphics[width=5cm]{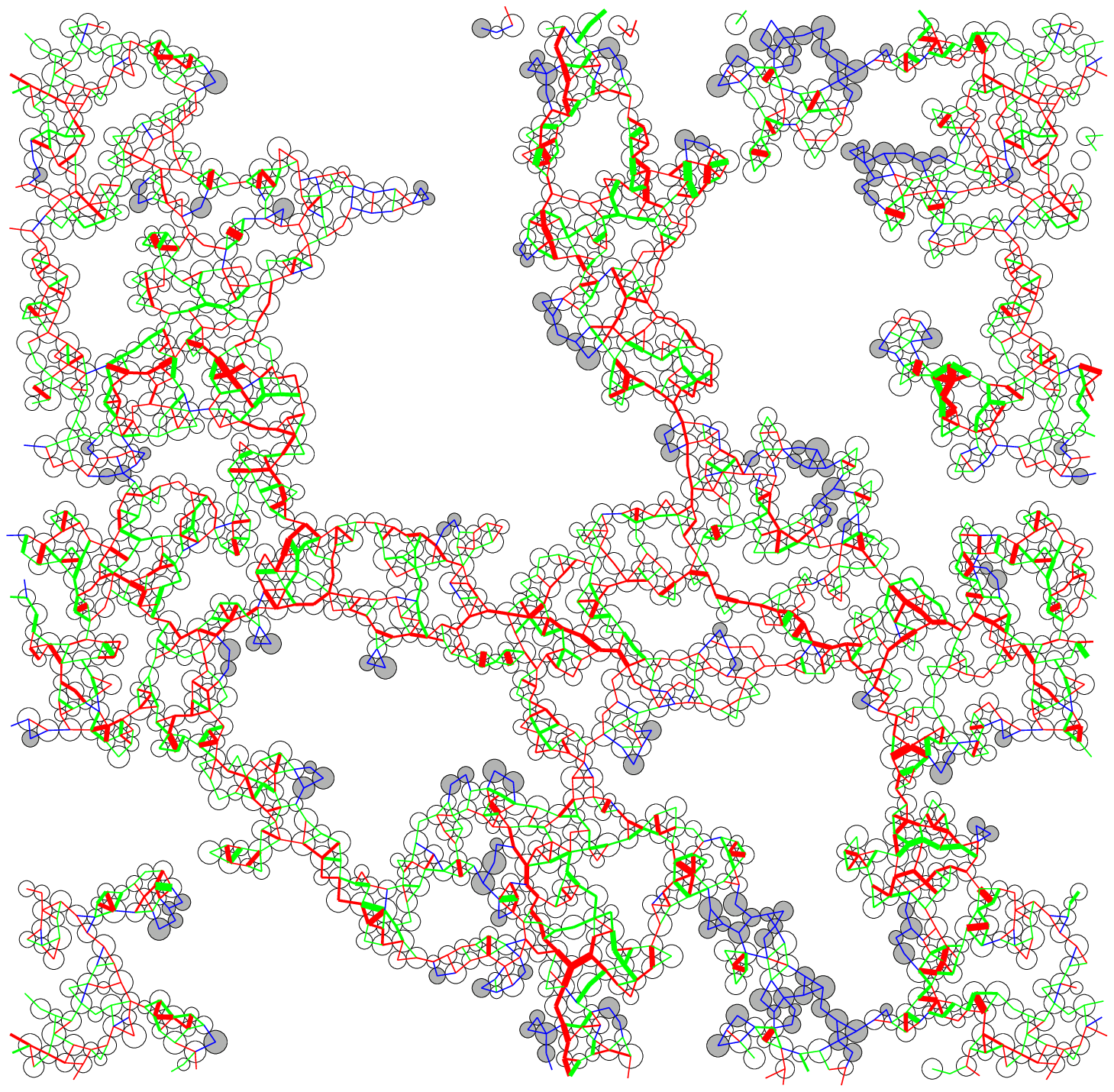}
\hfil
\includegraphics[width=5cm]{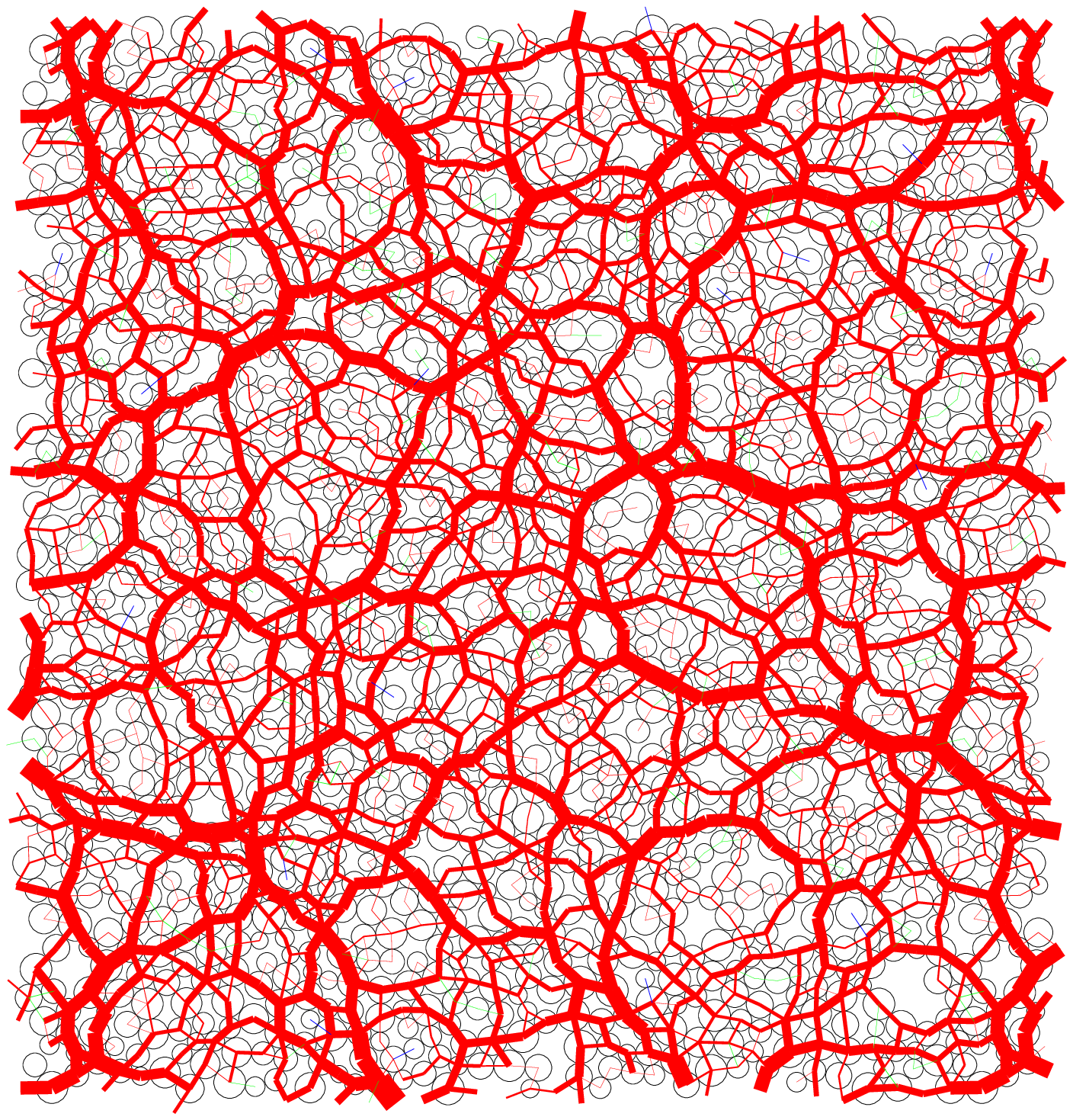}
\caption{\label{fig:lachedenseco} Aspect de la configuration pour
 $P^*=10^{-2}$, à gauche et pour $P^*=13$ (et $\kappa\simeq 7500$), à droite, d'un échantillon 2D de disques
cohésifs~\cite{Gilabert08}. }
\end{figure}

\begin{figure}[!htb]
\centering
\includegraphics[width=6.5cm]{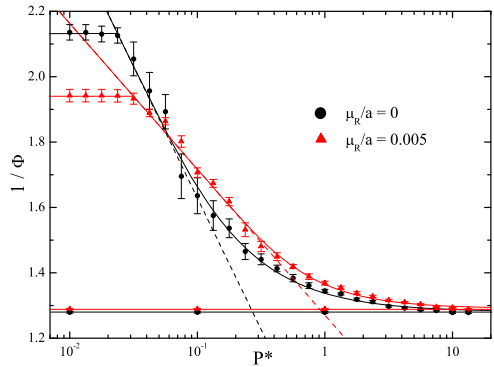}
\caption{\label{fig:conso} Courbe de consolidation en compression
isotrope et effet de la décompression dans un système modèle cohésif 2D,
($a$ est ici le diamètre maximal d'une distribution uniforme,
le minimum étant $a/2$)~: variations de $1/\Phi$ (ou
de l'indice des vides $-1+1/\Phi$) en fonction de $\ln P^*$~: une
faible valeur de $\mu_R/a$ suffit à en modifier la pente dans le
régime linéaire. Noter la faiblesse de la déformation à la décharge (partie inférieure
du graphe).
Résultats tirés de~\cite{Gilabert08}.}
\end{figure}

La compression isotrope entraîne très peu de déformation irréversible
avec des grains sans cohésion~\cite{iviso2}, en dépit de certaines
évolutions internes (une coordinence $z$ initialement élevée
peut diminuer dans un cycle de compression).
La situation est tout autre avec des grains cohésifs, car une
augmentation de pression modifie le rapport $\eta = 1/P^*$, défini
en~\eqref{eq:defeta}, entre la cohésion, capable de stabiliser des
structures lâches, et la pression de confinement. L'application
d'un confinement modifie alors considérablement l'état du système,
la compression sous chargement isotrope (ou sur un trajet de
chargement tel que le rapport des contraintes reste constant)
conduit à une déformation plastique très importante et définit la courbe de consolidation
(figure~\ref{fig:conso}). Cette évolution de $1/\Phi$ (ou de l'indice des vides) avec
le logarithme de la pression est un résultat classique pour les poudres et les sols cohésifs~\cite{BiHi93},
et on la mesure en laboratoire dans la limite
quasi-statique. À faible $P^*$, l'approche de cette limite, lors de la compression, est à évaluer
avec le paramètre $I_a$ de~\eqref{eq:defIa}~\cite{Gilabert08}.
La courbe de consolidation est sensible à
la résistance au roulement, même très faible,
$\mu_R/d = 10^{-2}$. La résistance au roulement joue un rôle
micromécanique important dans les structures cohésives
lâches \cite{Gilabert07,Gilabert08}, car elle permet la transmission
d'efforts par des chaînes de grains, alors dotées d'un raideur en
flexion.
\subsection{Matériaux non cohésifs, déformation quasi-statique}
Nous prenons l'exemple de la simulation du comportement d'un
assemblage de billes identiques non cohésives dans une compression
triaxiale de révolution, essai classique en mécanique des
sols~\cite{BiHi93}, souvent simulé par éléments discrets avec des
matériaux modèles~\cite{TH00,SUFL04}. Le taux de déformation
axiale $\dot\epsilon_a$ est alors imposé, et $P$, dans la
définition de $I$ et de $\kappa$, désigne la pression isotrope
initiale, à laquelle restent égales les deux contraintes
principales $\sigma_2$ et $\sigma_3$, tandis que $\sigma_1 =P+ q$
augmente avec $\epsilon_1 = \epsilon_a$.
Il s'agit normalement d'une sollicitation quasi-statique. Alors
que le nombre d'inertie dans un essai triaxial de laboratoire sur
un sable prend souvent des valeurs de l'ordre de $10^{-9}$, les
simulations les plus soigneuses ne descendent guère en dessous de
$I=10^{-5}$. Pour que l'essai numérique soit aussi lent que
l'essai de laboratoire il faudrait donc de l'ordre d'une année de
calcul au lieu d'une heure. Les effets dynamiques peuvent donc se
manifester bien davantage dans le calcul numérique, et il faut
s'assurer qu'ils n'influent pas trop sur les comportements
mécaniques et les mécanismes rhéophysiques observés.
\begin{figure}[!htb]
\centering
\includegraphics[width=.47\textwidth]{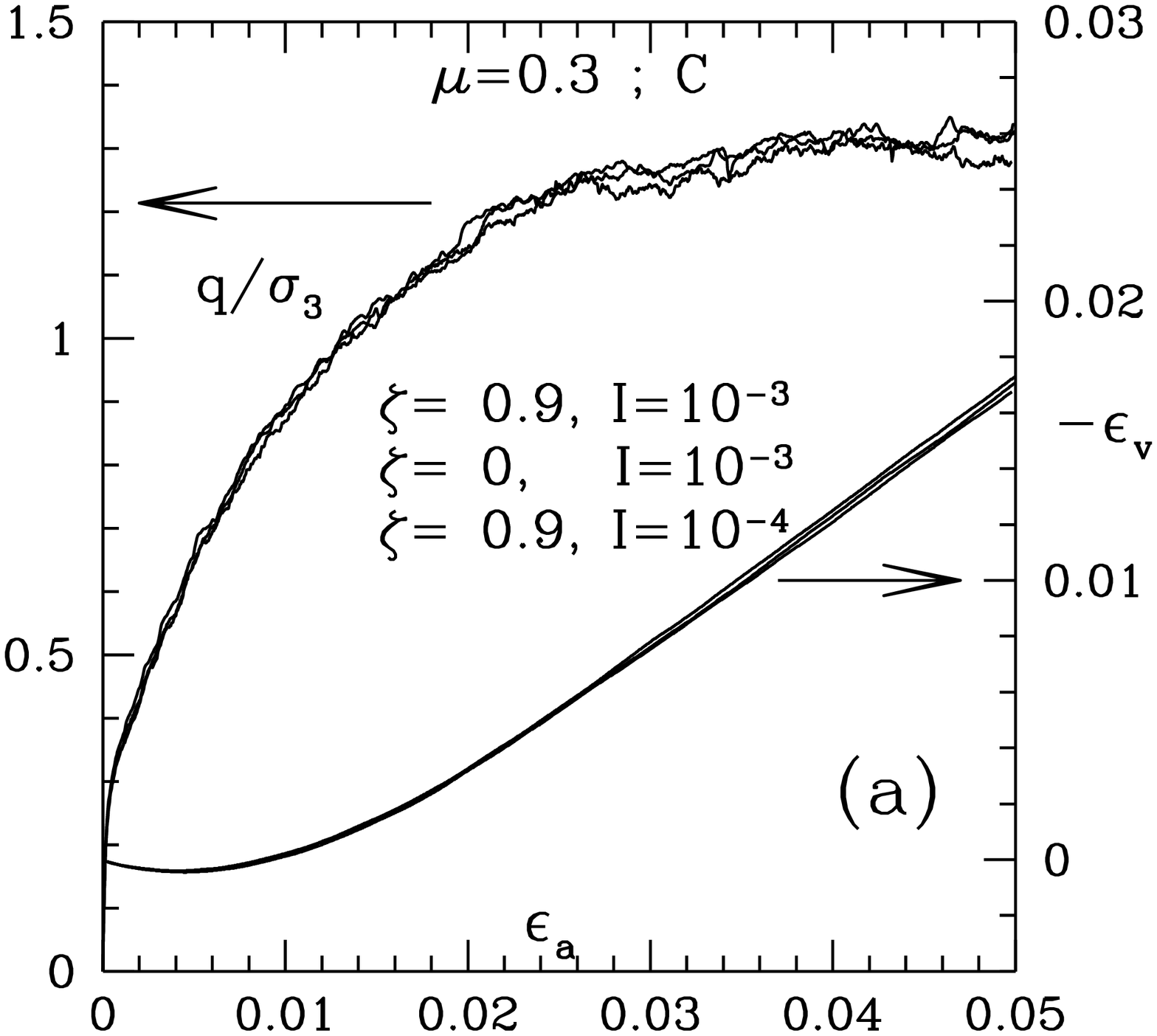}
\hfill
\includegraphics[width=.47\textwidth]{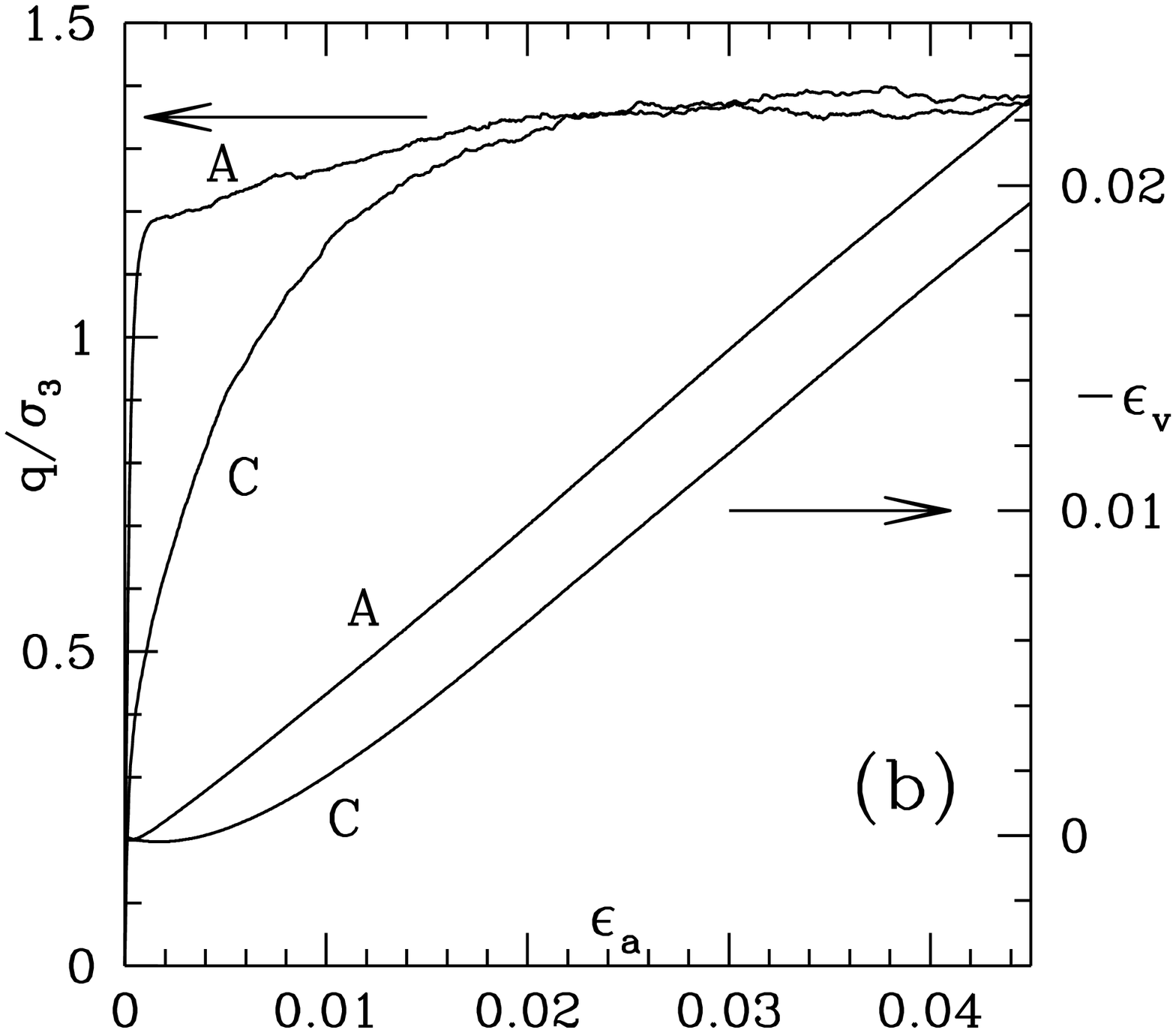}
\caption{\label{fig:monteepic1} 
Déviateur des contraintes $q$,
normalisé par $P$ (axe de gauche) et déformation volumique
$\epsilon_v$ (axe de droite) fonctions de la déformation axiale
dans la compression triaxiale d'assemblages de billes
(Hertz-Mindlin, $\mu=0$,3, $\mu_R=0$, $\kappa = 8400$). On compare~: 
(a) pour le même état de départ, différentes valeurs de $I$ (Éq.~\ref{eq:defI}) et $\zeta$ (cf. §\ref{sec:visq})~;
(b) deux états initiaux, A et C,  de même densité, mais de nombre de coordination
différent. }
\end{figure}
La figure~\ref{fig:monteepic1} représente la réponse au test
triaxial, exprimée selon l'usage avec le déviateur des contraintes
$q$ et la déformation volumique $\epsilon_v$ fonctions de la
déformation axiale $\epsilon_a$, d'échantillons numériques de
N=4000 billes, entre l'état isotrope de départ et le maximum de
déviateur.

La figure~\ref{fig:monteepic1}a montre que, pour le
cas considéré, celui d'un système dense, un nombre d'inertie
inférieur à $10^{-3}$ peut suffire à approcher la limite
quasi-statique de la courbe contrainte-déformation. Les courbes
macroscopiques apparaissent alors indépendantes du niveau de dissipation
visqueuse (on peut même prendre $\zeta=0$) et du nombre d'inertie $I$.
Toutefois, au niveau des grains, il
subsiste des écarts à l'équilibre, qui dépendent de $\zeta$ et
tendent à se réduire si $I$ décroît. Il en résulte
un certain fluage lorsqu'on arrête d'imposer la valeur de
$\dot\epsilon_a$ pour maintenir $q$ constant. Le système se
stabilise alors finalement dans un état d'équilibre (avec les critères
discutés au chapitre 1), après que $\epsilon_a$ et $\epsilon_v$ ont subi, typiquement
une augmentation d'ordre $10^{-3}$ ou $10^{-4}$ dans le cas
considéré ici. Ces déformations différées dépendent de divers paramètres
dynamiques ($\zeta$, $I$, pilotage des conditions aux limites...) mais
diminuent progressivement à mesure que $I$ décroît.

\begin{figure}[!htb]
\centering
\includegraphics[angle=270,width=6cm]{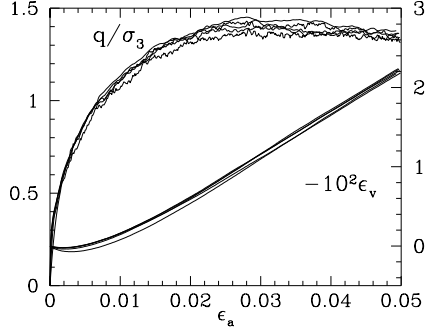}
\caption{\label{fig:dessPQ} Effet de $\kappa$
sur les courbes $q(\epsilon_a)$ et $\epsilon_v(\epsilon_a)$ pour le système C de
la figure~\ref{fig:monteepic1} (même représentation). Les états initiaux
modélisent ici des billes de verre sous confinement croissant,
$P=10$, 32, 100, 320 et 1000~kPa, soit $\kappa$ décroissant de 39000 à 1800.}
\end{figure}

\begin{figure}[!htb]
\centering
\includegraphics[width=.47\textwidth]{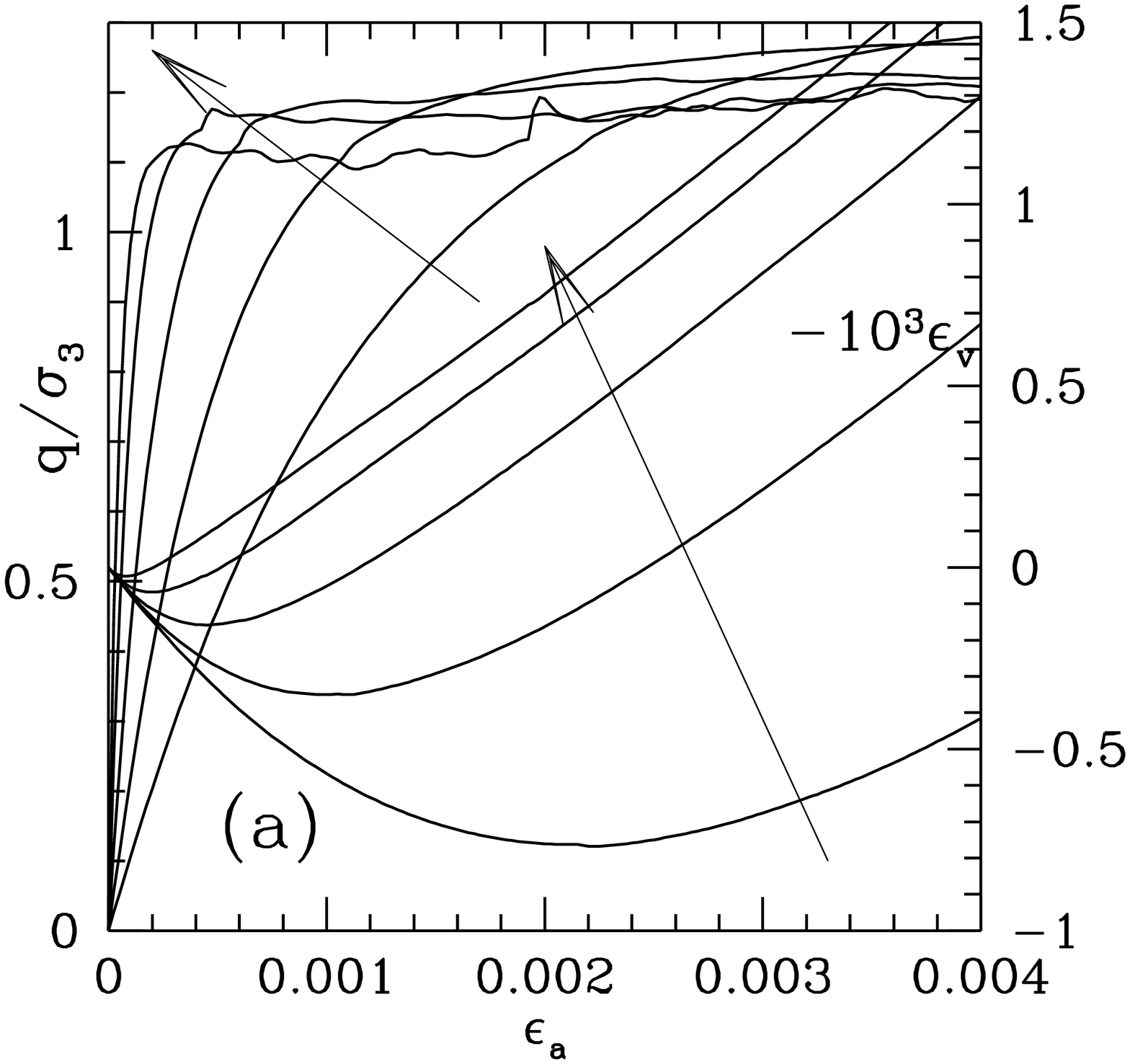}
\hfill
\includegraphics[width=.47\textwidth]{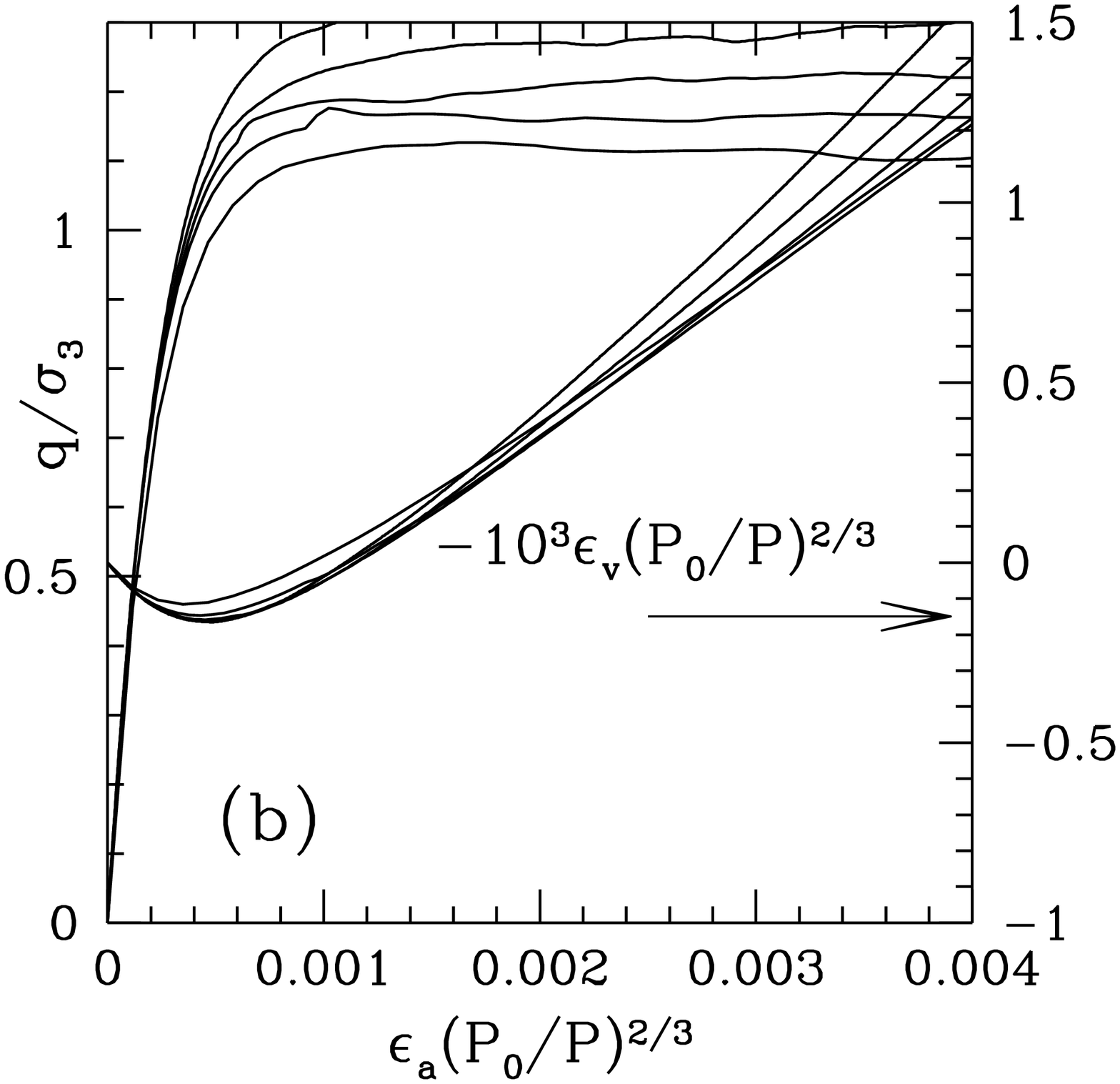}
\caption{\label{fig:dessPP}
(a) : analogue de la figure~\ref{fig:dessPQ} pour l'assemblage A (mêmes valeurs de $P$).
Les flèches classent les 5 courbes $q(\epsilon_a)/P$
et $\epsilon_v(\epsilon_a)$ selon $\kappa$ croissant (ou $P$ décroissante). 
(b) : les mêmes résultats avec les déformations à l'échelle $(P/P_0)^{2/3}\propto \kappa^{-1}$ ($P_0=100$~kPa).
}
\end{figure}

La figure~\ref{fig:monteepic1}b illustre
l'importance de la géométrie de la configuration initiale pour le
comportement mécanique (comme annoncé au
§~\ref{sec:geom}). Au-delà de la seule compacité, très proche
pour A et C de la valeur maximale $\Phi^* \simeq 0,64$ pour les
assemblages désordonnées de billes, on voit que le nombre de
coordination initial $z$ détermine la forme de la courbe de déviateur, qui est beaucoup
plus raide pour A ($z\simeq 6$) que pour C ($z\simeq 4$). De plus, ces deux assemblages qui
diffèrent par leur coordinence initiale ne présentent pas du tout la même sensibilité
au niveau de raideur $\kappa$. Le comportement du
matériau C, aux échelles de déformation considérées,
est indépendant de $\kappa$ (si on l'exprime avec des grandeurs
sans dimension comme $q/P$, $\epsilon_v$), comme le montre la
figure~\ref{fig:dessPQ}. Les courbes relatives à différentes
valeurs de $\kappa$ pour le matériau A diffèrent nettement (figure~\ref{fig:dessPP}a), mais
tendent à se superposer dans un intervalle de déviateur assez étendu (figure~\ref{fig:dessPP}b) si on adopte
l'échelle $\kappa^{-1}$ pour les déformations, en représentant
$q/P$ et $\kappa\epsilon_v$  comme fonctions de $\kappa\epsilon_a$.

L'explication de ces différents comportements est donnée au
chapitre 1, où sont distingués deux types de déformations, celles
qui proviennent de la déformation des contacts (type I), et celles
qui résultent de réarrangements du réseau, suite à des
instabilités microscopiques (type II). Les systèmes bien
coordonnés, comme ici l'assemblage A, possèdent un réseau de
contact initial capable de résister à l'application d'un important
déviateur, et donc présentent une réponse de type I, qui, bien que
non élastique~\cite{Roux05,iviso3} en raison de la mobilisation du
frottement intergranulaire, est telle que la déformation soit
inversement proportionnelle au niveau de raideur. La rupture
de la structure des contacts survient à l'approche du maximum de déviateur,
et on observe alors des déformations de type II.
Lorsque, comme
dans le cas de l'assemblage C, le réseau des contacts est mal
coordonné et cède facilement, la déformation de type II domine pour
des déviateurs beaucoup plus faibles, et
c'est la géométrie du paquet de grains en contact qui régit
l'amplitude des déformations, devenues insensibles à la raideur
$\kappa$. L'amplitude des déformations est alors liée aux épaisseurs des
interstices entre grains voisins, dont la fermeture crée de nouveaux
contacts et stabilise une nouvelle configuration du réseau après que
l'incrément de contrainte (ou de déformation) imposé
a conduit à l'instabilité et à la rupture de la précédente.
En posant que les déformations d'un réseau stable (type I), sont
inversement proportionnelles au niveau de raideur, nous admettons que
sa géométrie n'y est pas sensible, ce qui est la définition de la
limite des grains rigides. On voit donc
que {\em la limite des grains rigides correspond à la séparation d'échelle entre les déformations
de type I  (à l'échelle $\kappa^{-1}$) et les déformations de type II (dont l'échelle est fixée par les interstices).}

Le domaine élastique, accessible à l'expérience, est strictement inclus
au régime de déformation de type I, plus étendu.
Les déformations de type I, bien que d'ordre $\kappa^{-1}$, restent donc mesurables~\cite{iviso3}, et on doit prendre
la bonne valeur de $\kappa$ pour les modéliser correctement en simulation. C'est dans ce régime que la déformation
peut aussi dépendre de $K_T/K_N$ (ou du coefficient de Poisson $\nu$ du matériau constituant les grains). Cependant,
les résultats de~\cite{Gael2} sur les assemblages de disques
(avec un intervalle de déviateur important en régime I) montrent que le comportement mécanique est
à peu près insensible à ce paramètre s'il dépasse $1/2$. Une curiosité~: le choix (plutôt injustifié pour des
grains solides ordinaires) de $K_T>K_N$ peut donner des matériaux à coefficient de
Poisson négatif~\cite{Gaspar08}. Quant aux raffinements des lois
\eqref{eq:hertz}-\eqref{eq:mindlin} proposés dans \cite{THRA88}, on a pu vérifier
\cite{iviso3} qu'ils n'influent que très peu sur les propriétés élastiques des assemblages de sphères.

Quand on prépare un échantillon numérique dense en supprimant le frottement dans la phase d'assemblage (chapitre 8),
on obtient une configuration de coordinence élevée, analogue à l'état A pris en exemple ici (figure~\ref{fig:dessPP}).
Dans les résultats expérimentaux sur les sables, la croissance de $q$ avec $\epsilon_a$ est très souvent
beaucoup plus lente que celle que l'on obtient alors dans la simulation.
Il est alors tentant d'attribuer aux raideurs choisies pour les calculs une valeur trop
faible (par exemple $\kappa=100$). Ce faisant, on décale effectivement le maximum de
déviateur vers des déformations plus grandes (figure~\ref{fig:dessPP}a), mais on
commet l'erreur d'adopter un modèle physiquement non pertinent pour les déformations. À cet égard il est préférable
de choisir l'état initial noté C  dans les exemples précédents (figure~\ref{fig:monteepic1}b).

Les résultats rappelés ici (figure~\ref{fig:monteepic1}b) indiquent aussi que 
le rapport $q_{\,\text{max}}/P$ du maximum de
déviateur (le << pic >>)  à la contrainte latérale  ne dépend pas de
$\kappa$ (ici $\ge 1800$) et est essentiellement sensible, pour un
matériau donné, à la compacité initiale $\Phi$ de l'échantillon (avec ici
$\Phi_C\simeq\Phi_A=\Phi^*$), conformément aux résultats de la mécanique des
sols \cite{BiHi93}. L'angle de frottement interne $\varphi$ est fonction croissante
du coefficient de frottement intergranulaire, comme l'indique
la table~\ref{tab:phiphi}, qui fournit également
les valeurs  de la dilatance $\dila$ lorsque $q=q_{\,\text{max}}$. Les données
de la table~\ref{tab:phiphi} sont donc définies par~:
\be
\frac{q_{\,\text{max}}}{P} = \frac{2\sin\varphi}{1-\sin\varphi}\ ;\ \ \ \ \ \ \ \ \
\dila = -\left(\frac{d\epsilon_v}{d\epsilon_a}\right)_{q=q_{\,\text{max}}}.
\label{eq:dilatpic}
\ee
\begin{table}
\centering
\begin{tabular}{|c||c|c|c|c|}  \cline{1-5}
$\mu$ &0&$0.3$&$0.5$&$1$ \\
\hline
$\varphi_c$ ($\degr$)& 0&$16.7$ &$26.6$   & $45$ \\
\hline \hline
$\varphi$ ($\degr$) & $5.5\pm 0.5$&$24\pm 1$ & $26\pm 1$ & $28 \pm 1$  \\
\hline
$\dila$ &$0\pm 0.01$ & $0.52\pm 0.01$ &$0.60\pm 0.01$ &$0.71\pm 0.01$ \\
\hline
\end{tabular}
\caption{\label{tab:phiphi} Angle de frottement interne
$\varphi$ et dilatance $\dila$ pour $q=q_{\,\text{max}}$, fonctions de
$\mu= \tg \varphi_c$, pour des assemblages
de sphères monodisperses ($\mu_R=0$). 
Valeurs issues de simulations de compressions triaxiales à partir d'états initiaux isotropes,
avec $\Phi\simeq\Phi^*$~\cite{Emam06}.}
\end{table}
Ces résultats sont relatifs à la compacité
maximale $\Phi^*$ des configurations désordonnées de 
sphères d'une seule taille.  
Il est traditionnel de relier $\varphi$ à $\varphi_c$ et à $\dila$, le frottement intergranulaire
et la dilatance apparaissant comme les origines du frottement interne~\cite{PWR62}
(voir aussi~\cite{TER06}). Cependant, il faut noter (table~\ref{tab:phiphi}) que les assemblages de grains
non frottants  sont dotés de frottement interne mais dépourvus de dilatance~\cite{Peyneau08}. C'est la microstructure
des assemblages granulaires, qui, en développant, sous l'effet de la déformation, une certaine anisotropie~\cite{CCL97},
permet au matériau de supporter des contraintes déviatoires croissantes~\cite{RB89,Radjai04,ARPS07,PR08b}. Il s'agit en
général d'un phénomène de réarrangement géométrique, associé à des déformations de type II.
On a pu noter que l'introduction d'un faible frottement de roulement pouvait sensiblement augmenter le
niveau de déviateur maximal~\cite{GaelPG09,TER06}. Avec des particules
ellipsoïdales plutôt que sphériques, le frottement interne augmente aussi
très notablement~\cite{AnKu04}.

\subsection{L'état critique} \label{sec:etatcritique}

Les simulations confirment~\cite{TH00,Radjai04,RK04} la théorie
de l'état critique~\cite{BiHi93}~: pour de grandes
déformations, l'état interne du matériau s'approche d'un certain
attracteur, acquiert une certaine \emph{structure d'écoulement},
indépendamment de l'état initial.
Une fois cet \emph{état critique} atteint, la déformation plastique peut
augmenter indéfiniment, de façon monotone, à déviateur de
contraintes et densité constants.
L'état critique correspond à la limite quasi-statique des
écoulements granulaires, objets du §~\ref{sec:ecoulement}. Lorsque
la déformation se localise dans une ou plusieurs bandes de
cisaillement, le matériau se trouve dans l'état critique à
l'intérieur de ces bandes~\cite{Fazekas07}.
\begin{figure}[!htb]
\centering
\includegraphics[angle=270,width=.48\textwidth]{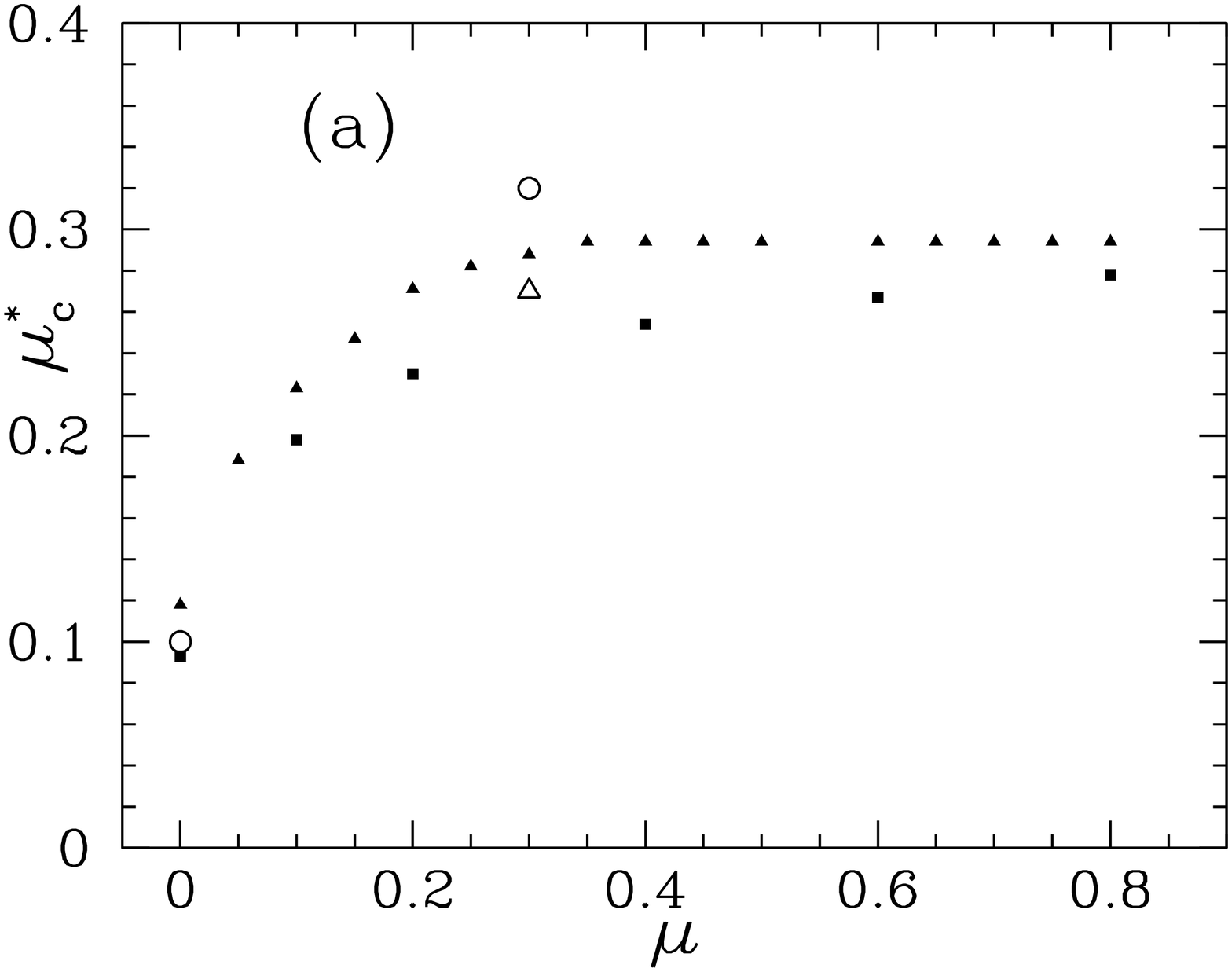}
\hfil
\includegraphics[angle=270,width=.48\textwidth]{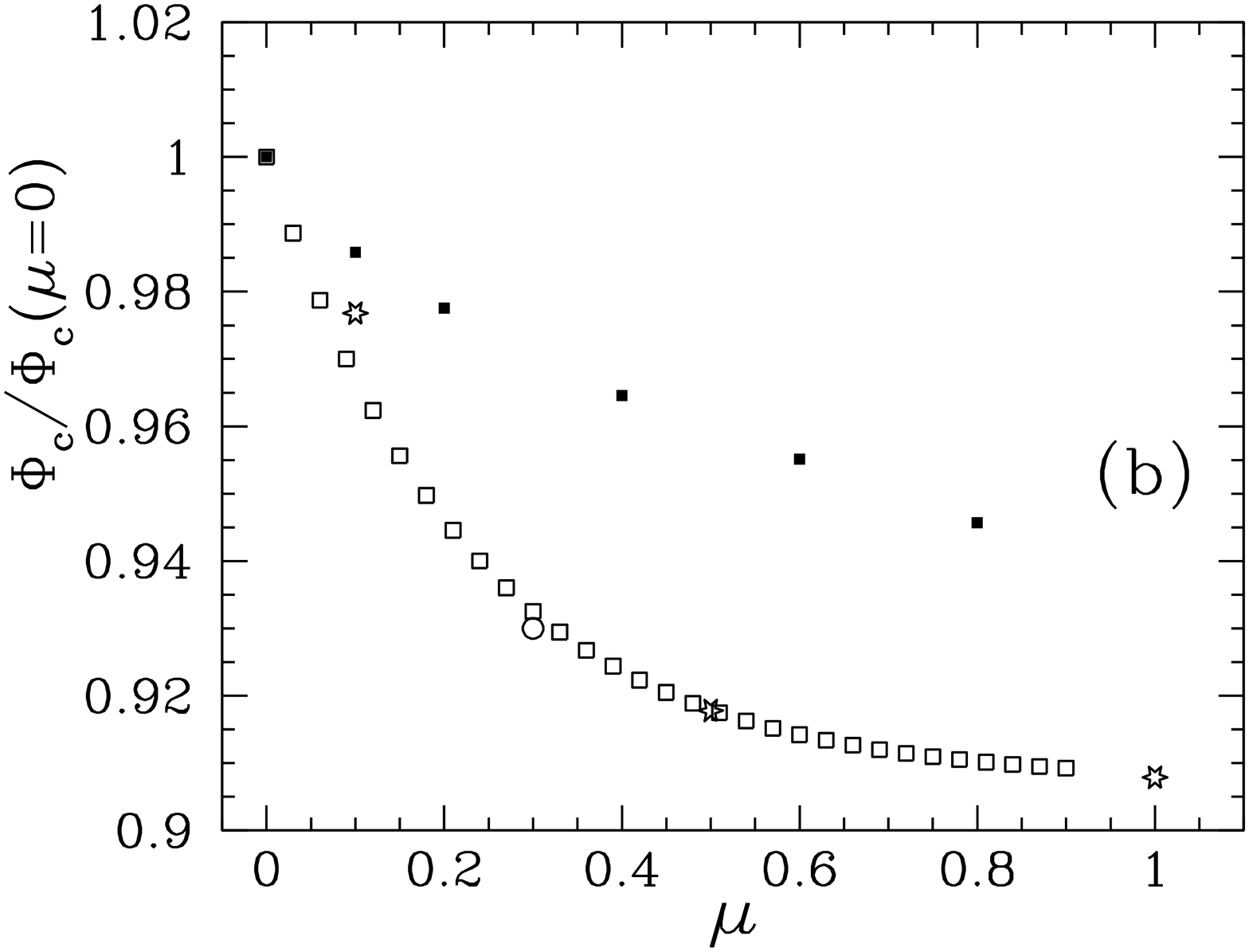}
\caption{\label{fig:critiq} Frottement interne à l'état critique,
$\mu^*_c$ (a), et rapport $\Phi_c/\Phi^*$ de la compacité critique
à la compacité maximale des assemblages aléatoires (b)
($\Phi^*\simeq 0,639$ en 3D, et $\Phi^*\simeq 0,85$ en 2D). Les
résultats proviennent de différentes études numérique publiées,
certaines pour des grains d'une seule taille, d'autres
introduisant une certaine polydispersité (diamètres uniformément
distribués entre $(1-x)d$ et $(1+x)d$~, pour des systèmes 2D
(symboles pleins)~: \cite{Estrada08} (carrés, $x=0,11$)~;
\cite{TER06} (triangles, $x=0,6$)~; et en 3D (symboles ouverts)~:
\cite{RoCh05} et \cite{Peyneau08} (cercles, $x=0$)~; \cite{TH00}
(triangle, $x\simeq 0,6$)~; \cite{Fazekas07} (carrés, $x=0$)~;
\cite{Campbell05} (étoiles, $x=0$).}
\end{figure}
Relativement bien documenté en simulation en raison de son
indépendance vis-à-vis des questions délicates d'assemblage et de
conditions initiales, l'état critique est indifféremment étudié
par des méthodes numériques qui prennent ou non en compte la
déformabilité des grains, et est insensible au niveau de raideur
$\kappa$ s'il est assez élevé (supérieur à quelques milliers). Il
est caractérisé par le frottement interne et la densité critiques,
qui ne dépendent alors que de facteurs géométriques et du
coefficient de frottement $\mu$ (si $\mu_R=0$). La figure~\ref{fig:critiq}
donne le frottement interne et la densité à l'état critique\footnote{
Les données de la 
figure~\ref{fig:critiq} correspondent à l'état critique en cisaillement simple pour les systèmes 2D et
les systèmes 3D de~\cite{Campbell05}, \cite{Peyneau08} et~\cite{Fazekas07}. 
Celles de~\cite{TH00} et de ~\cite {RoCh05} ont été obtenues en compression triaxiale. L'angle de frottement interne peut être
différent dans ces deux états -- c'est la cas en particulier pour la valeur en $\mu=0$~\cite{PR08b}.} en
fonction de $\mu$ pour des systèmes de disques (2D) ou de sphères
sans résistance au roulement.

L'état critique est très sensible au frottement de roulement~\cite{Estrada08}.
La référence \cite{thVoivret} en donne une étude pour un matériau (2D) de
grande polydispersité. Il y est montré en particulier que l'angle de frottement interne
ne dépend pas sensiblement de la granulométrie.

\section{Écoulements} \label{sec:ecoulement}

On vient de voir que l'état critique des assemblages granulaires
non cohésifs, dans lequel le matériau se déforme continuellement,
dans la limite quasi-statique, en maintenant une structure interne
constante, dépend exclusivement, parmi les différents paramètres
de contrôle introduits au §~\ref{sec:etatcritique}, du coefficient
de frottement (pour des $\kappa$ assez grands). Nous nous
proposons maintenant d'étudier les écoulements stationnaires
uniformes~\cite{Chevoir08d}, dans lesquels l'état du système a
ceci de commun avec l'état critique qu'il est maintenu constant,
indépendant de l'état initial, et ceci
de différent que l'on n'est plus dans la limite des faibles
nombres d'inertie.

La loi de comportement va alors s'exprimer à travers les
dépendances des variables caractérisant l'état stationnaire du
système en fonction des paramètres de contrôle sans dimension
précédemment introduits. On commence par le cas de matériaux
granulaires non cohésifs dans la géométrie la plus simple, celle
du cisaillement plan sans gravité où la distribution des
contraintes est
homogène~\cite{Dacruz04a,Dacruz05,Rognon06a,Rognon08a}. On discute
alors les influences respectives de l'état de cisaillement (à
savoir le nombre inertiel $I$) et des caractéristiques mécaniques
des contacts (à l'exclusion toutefois du frottement de roulement)
sur la loi de comportement. On aborde ensuite l'influence du
gradient de contraintes~\cite{Koval09a}, et enfin l'influence de
la cohésion~\cite{Rognon06a,Rognon08a}, à travers le nombre
$\eta$.

\subsection{Cisaillement homogène} \label{sec:cishom}

Dans cette configuration générique, le système occupe une cellule
de longueur $L$ et de hauteur $H$, avec des conditions aux limites
périodiques dans la direction du cisaillement (voir chapitre 6).
On impose la pression latérale $P$~\footnote{On constate, comme
dans d'autres géométries, que les contraintes normales sont
approximativement égales.}, plutôt que la compacité~\footnote{Ceci
est approprié pour discuter de situations réelles puisque les
conditions expérimentales imposent habituellement le niveau de
contrainte plutôt que la compacité. On a montré que les
simulations à pression contrôlée sont en très bon accord avec
celles à compacité contrôlée~\cite{Dacruz04b, Lemaitre08}.}, de
sorte que la hauteur $H$ n'est pas absolument fixée. On cherche
par ailleurs à imposer le taux de cisaillement $\dot\epsilon$.
Deux types de condition aux limites ont été testés dans la
direction perpendiculaire au cisaillement (voir chapitre 6). La
première méthode consiste à placer le matériau entre deux parois
rugueuses et parallèles. Pour imposer le cisaillement, une des
parois est fixe et l'autre se déplace à la vitesse $V$. La seconde
méthode, dite de \emph{Lees-Edwards} (on parle aussi de
\emph{conditions aux limites bipériodiques})~\cite{Radjai02},
consiste à répéter périodiquement la cellule dans la direction $y$
orthogonale à l'écoulement. Pour imposer le cisaillement, les
cellules inférieure et supérieure sont animées d'une vitesse $\pm
V$~\footnote{La vitesse $V$ de la paroi ou des cellules images
s'adapte aux fluctuations de hauteur pour maintenir un taux de
cisaillement constant.}. Le contrôle de la pression $P$ est assuré
en permettant la dilatation de la cellule selon $y$ au cours du
temps.

On évoque dans la suite les résultats obtenus pour des systèmes
bidimensionnels (les contraintes sont alors homogènes à une force
par unité de longueur), simulés par la méthode de dynamique
moléculaire~\cite{Dacruz04a,Dacruz05,Rognon06a,Rognon08a}. On en
tire des enseignements généraux, vérifiés dans d'autres
simulations (assemblages de sphères~\cite{Peyneau08}, méthode de
dynamique des contacts~\cite{Lois05}). Sauf mention contraire, le
matériau granulaire est une collection dense de $N$ disques
dissipatifs de diamètre moyen $d$ et masse moyenne $m$. Une petite
polydispersité de $\pm 20 \%$ est introduite pour empêcher la
cristallisation.

Comme on souhaite étudier la rhéologie intrinsèque
du matériau et non les effets de paroi, les parois sont
systématiquement assez rugueuses afin d'éviter une localisation du
cisaillement à leur voisinage. Elles sont
constituées de grains jointifs de mêmes caractéristiques que
les grains cisaillés, assemblés pour former un seul solide.

Les propriétés mécaniques des contacts sont décrites par les trois
paramètres mécaniques indépendants introduits au §~\ref{sec:lois}: 
le coefficient de frottement $\mu$, le
coefficient de restitution dans les collisions binaires $e$ et la 
raideur $K_N$ ($K_T$ est du même ordre de grandeur
que $K_N$~\cite{Johnson85}, et comme il a une très petite
influence sur les résultats~\cite{Silbert01,Campbell02}, il a été
fixé à $K_N/2$ dans tous les calculs). Sauf mention contraire,
les résultats rapportés ici correspondent à $\mu=0,4$, $e=0,1$ et $\kappa=10^4$. 
Ceci constitue le matériau de
référence pour la suite de la discussion. La longueur $L$ est
choisie supérieure ou égale à 40 grains pour ne pas avoir d'effet
de taille~\cite{Prochnow02}. L'épaisseur $H$ varie entre 20 et 100
grains, de sorte que $N$ varie entre $1000$ et $5000$.

Le système évolue vers un \'etat de cisaillement stationnaire et
uniforme, ind\'ependant du mode de pr\'eparation (\'etat l\^ache
par d\'ep\^{o}t al\'eatoire, ou dense par compaction cyclique sans
frottement), caractérisé par des profils stationnaires
de compacité, vitesse et contrainte. On considère que
les distributions statistiques des quantités d'intérêt (structure,
vitesses, forces\dots) sont indépendantes de la direction du
cisaillement et du temps, de sorte que l'on moyenne à la fois dans
l'espace (selon la direction du cisaillement) et dans le temps sur
une durée $\Delta t \geq 10/\dot \gamma $.
\subsection{Loi de comportement}
\subsubsection{Influence du nombre inertiel}
\begin{figure}[!htbp]
\centering
\includegraphics*[width=10cm]{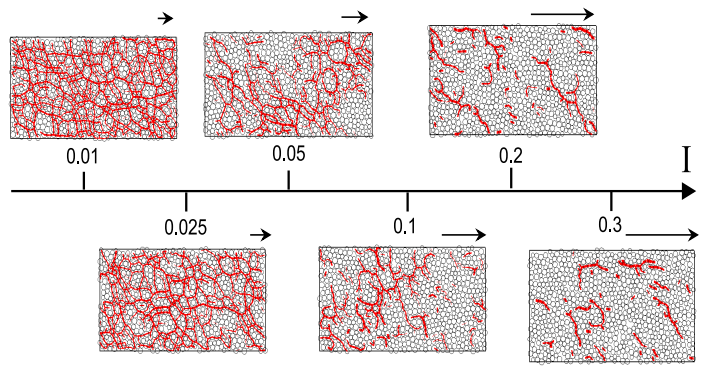}
\caption{\'Evolution du réseau de contact en cisaillement
plan homogène (taux de cisaillement et pression imposés) en
fonction du nombre inertiel $I$ : les traits représentent les
forces normales entre particules (d'après~\cite{Rognon06b}).
\label{Fig.Ecoul0}}
\end{figure}
\begin{figure}[!htbp]
\begin{center}
\includegraphics*[width=6cm]{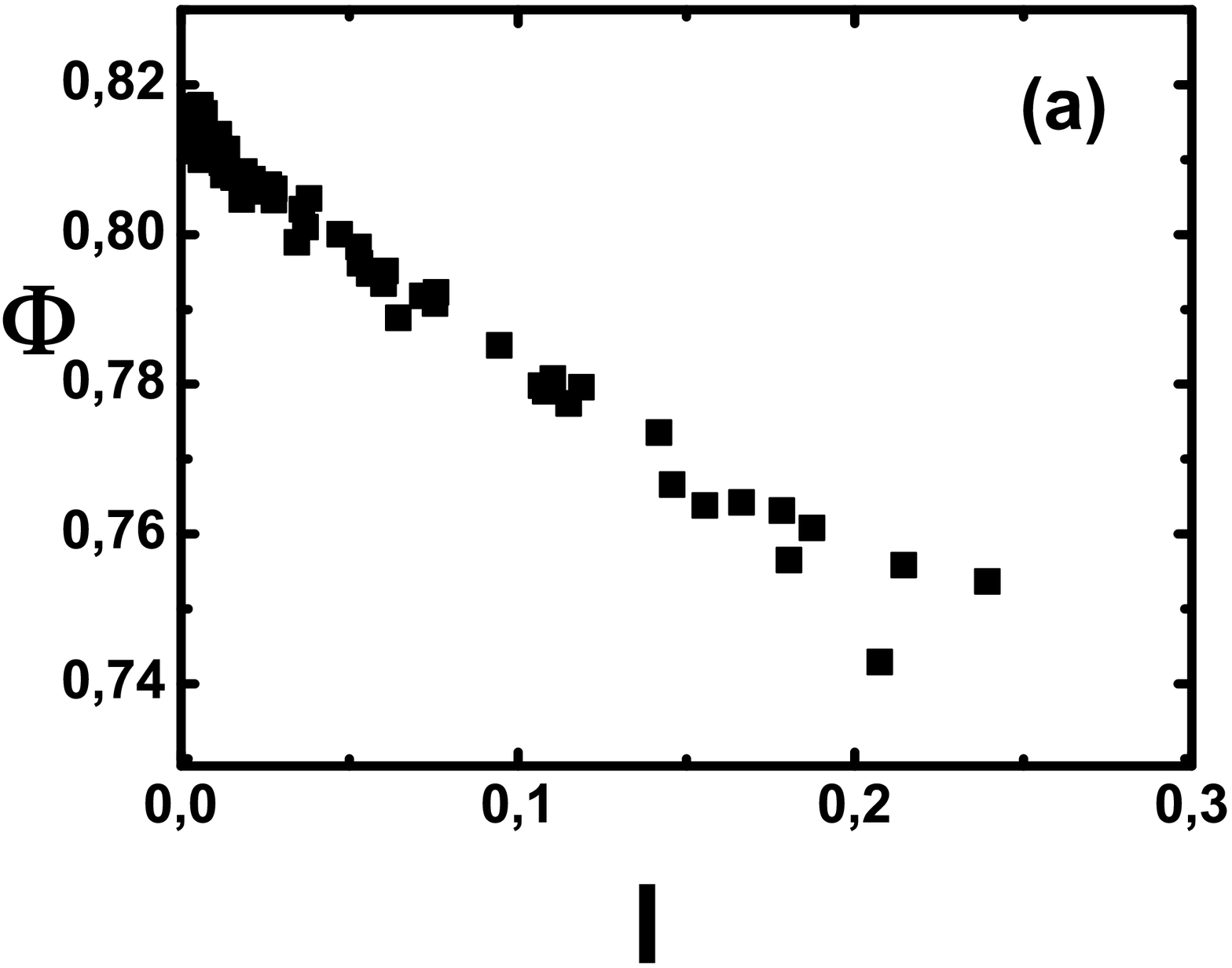}
\includegraphics*[width=5cm]{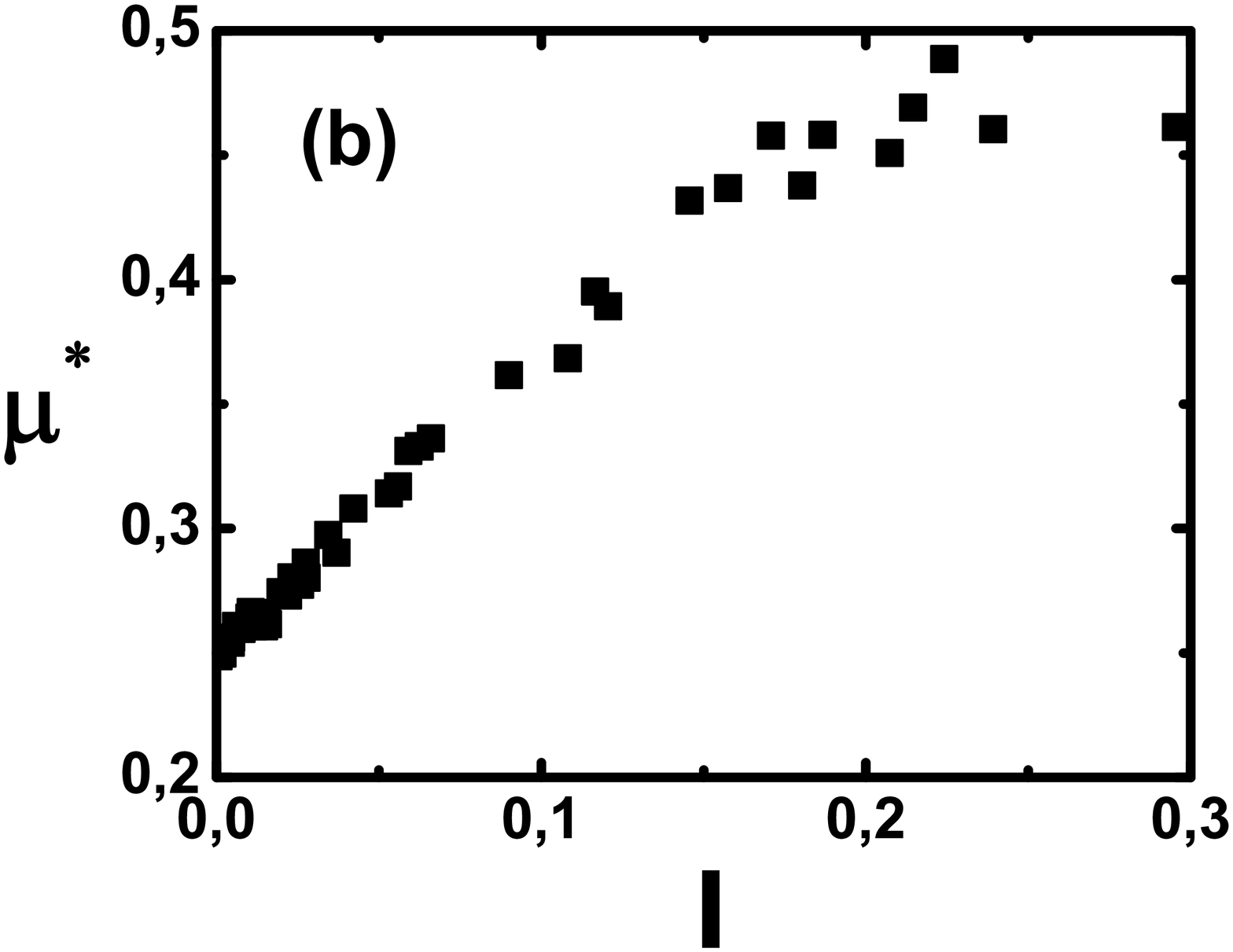}
\caption{\label{Fig.Ecoul1} \textit{(a) Loi de dilatance, (b) loi
de frottement ($\mu = 0,4$ - $e$ variable) -
d'après~\cite{Dacruz05}.}}
\end{center}
\end{figure}

L'analyse dimensionnelle a montr\'e que l'\'etat de cisaillement
est caract\'eris\'e par le nombre inertiel $I$, grandeur
fondamentale pour décrire le comportement rhéologique des
matériaux granulaires cisaillés. Une petite valeur de $I$
correspond à l'état critique décrit au §~\ref{sec:etatcritique}.
Inversement, une grande valeur de $I$ correspond au régime
collisionel, où les grains interagissent par des collisions
binaires que l'on peut considérer comme instantanées et non
corrélées~\cite{Goldhirsch03}. Entre ces deux régimes existe un
régime de cisaillement \emph{dense} pour lequel l'inertie des
grains n'est pas négligeable. Le matériau est au-dessus de son
seuil d'écoulement, dans un état liquide plutôt que gazeux. Les
mouvements de grains sont fortement corrélés, et l'hypothèse de
chaos moléculaire propre aux gaz dilués est alors en défaut. Il
existe un réseau de contact percolant, extrêmement fluctuant à la
fois dans l'espace et dans le temps. Plus le nombre inertiel
augmente, moins les chaînes de force sont longues et nombreuses
(figure~\ref{Fig.Ecoul0}). Dans les états de cisaillement
stationnaire et uniforme, à pression et taux de cisaillement
imposé, la compacité et la contrainte de cisaillement s'ajustent
en réponse au nombre inertiel imposé, à travers une \emph{loi de
dilatance} et une \emph{loi de frottement} qui décrivent les
variations approximativement linéaires des deux grandeurs sans
dimension, la compacité $\Phi$ et le frottement effectif $\mu^{*}
= S/P$, en fonction du nombre inertiel $I$
(Fig.~\ref{Fig.Ecoul1}):
\begin{equation}\label{eq:loisdilfrot}
\begin{aligned}
\Phi(I) \approx \Phi_{\text{max}} - a I,\\
\mu^*(I) \approx \mu _{\text{min}}^*  + bI.\\
\end{aligned}
\end{equation}
Les paramètres $\Phi_{\text{max}}$, $\mu _{\min }^*$, $a$ et $b$
dépendent des caractéristiques du matériau granulaire. Pour le
matériau de référence, $\Phi_{\text{max}} \simeq 0,82$, $\mu^*_{\text{min}}
\simeq 0,26$, $a \simeq 0,37$ et $b \simeq 1,1$.
Dans le régime quasi-statique ($I \le 10^{-2}$), le frottement
effectif minimal $\mu^*_{\text{min}}$ et la compacité maximale
$\Phi_{\text{max}}$ s'identifient au frottement interne $\tg \phi$ et à
la compacité $\Phi_{c}$ dans l'état critique~\cite{Radjai04,
Peyneau08}. Lorsque $I$ augmente, le milieu se dilate
l\'eg\`erement, la proportion de collisions augmente par rapport
aux contacts maintenus jusqu'\`a un r\'egime purement collisionnel
($I \ge 0,2$) dans lequel la dilatance devient plus importante et
le frottement effectif semble saturer. Ainsi, à taux de
cisaillement imposé, l'utilisation du nombre $I$ permet de traiter
les comportements liquides et solides et le passage entre les
deux, sans singularité.

Avec ces deux lois on peut réécrire la loi de comportement du
mat\'eriau dans le r\'egime interm\'ediaire sous forme
viscoplastique, avec un seuil d'\'ecoulement de Coulomb et des
contraintes visqueuses, dont l'origine microscopique se situe dans
l'organisation du r\'eseau de contacts.
Une formulation tensorielle de ces lois 
été proposée et appliquée au cas d'écoulements vraiment
tridimensionnels, pour lesquels la comparaison avec les résultats
expérimentaux s'est avérée très satisfaisante~\cite{Jop06}.
\subsubsection{Autre choix des paramètres de contrôle}
Certains auteurs~\cite{Campbell02} ont choisi de contrôler la
compacité $\Phi$ plutôt que la pression, et en conséquence, plutôt
que $I$ et $\kappa$, ils ont utilisé le couple de nombres sans
dimension $\Phi$ et $\alpha=\dot{\gamma}/\sqrt{K_N/m} =
I/\sqrt{\kappa}$, comme variables caractérisant l'état du matériau
granulaire en cisaillement stationnaire uniforme. $\alpha$ peut
être vu comme le rapport des temps de collision et de
cisaillement, ou des vitesses de cisaillement et du son (nombre de
Mach~\cite{Campbell02}). Les deux choix sont parfaitement
légitimes, puisque l'analyse dimensionnelle prédit soit $\mu^* =
f_1(I,\kappa)$ et $\Phi = f_2(I,\kappa)$, soit $\mu^* =
f_3(\Phi,\alpha)$ et $\kappa = f_4(\Phi,\alpha)$, les deux
résultats étant également valables. Le choix de $I$ et de $\kappa$
peut cependant sembler plus pratique pour plusieurs raisons.
D'abord, les variations des résultats avec $\Phi$, considéré comme
paramètre de contrôle, sont extrêmement rapides. Chaque matériau
possède une compacité critique $\Phi_c$, au-dessus de laquelle il
ne s'écoule plus, sauf si les contraintes sont si grandes que la
compression élastique des contacts compense la différence
$\Phi-\Phi_c$. Par contre, en-dessous de $\Phi_c$, un milieu
granulaire continuellement cisaillé est libre de s'écouler avec
une contrainte de cisaillement négligeable, sauf si la vitesse est
assez grande pour conduire à une pression significative. On doit
donc piloter $\Phi$ avec une grande précision pour observer des
niveaux de contrainte usuels. Ceci rend les comparaisons entre
différent systèmes granulaires difficiles, puisqu'il faudrait
connaître à l'avance la valeur de la compacité critique pour
chacun d'entre eux. De plus, la limite des grains rigides devient
singulière, puisque toutes les valeurs de $\Phi$ au-dessus de
$\Phi_c$ sont strictement interdites pour $\dot \gamma \neq 0$,
alors que toutes les propriétés des cisaillements avec $\Phi \le
\Phi_c$ sont proportionelles à une puissance de $\dot \gamma$ dans
cette limite~\cite{Lois05}. Ainsi, aucun changement de régime
n'est attendu par un changement de $\dot \gamma$, ce qui semble
contredire l'intuition sauf si l'on rappelle que la limite des
grains rigides, lorsque l'on augmente le taux de cisaillement,
finit par requérir des raideurs de contact déraisonnablement
élevées, du fait de la très grande valeur de la pression. A
l'inverse, si l'on utilise $I$ et $\kappa$ comme paramètres de
contrôle, aucune singularité n'intervient dans les résultats dans
les limites $I\to 0$ ou $\kappa\to \infty$, et des matériaux différents
devraient manifester des comportements similaires (à défaut d'être
quantitativement identiques) pour les mêmes valeurs de ces
paramètres (ce qui définit ainsi approximativement des \emph{états
correspondants}).

\subsubsection{Influence de l'élasticité des contacts}

En le faisant varier entre $40$ et $2,5 \cdot 10^5$, on observe
que $\kappa$ n'a pas d'influence sur la loi de comportement dès
qu'il dépasse $10^4$ (ou même $10^3$), ce qui est conforme à
l'observation dans l'état critique. Pour les petites valeurs
($40$), on observe une localisation près de la paroi en mouvement
dans les grands systèmes~\cite{Dacruz04a}, que l'on interprète
comme un effet de la décroissance de la longueur de corrélation du
champ de déformation quand les grains deviennent mous. On remarque
que $\kappa = 10^4$ correspond à $\tau_i/\tau_c = 10^2$. Puisque
$I = \tau_i/\tau_s$ est plus petit qu'environ $0,1$ dans le régime
dense, $\tau_s$ est au moins $1000$ fois plus grand que $\tau_c$
dans la limite des grains rigides. En conséquence, on peut bien
parler de limite des grains rigides dans le régime dense. De fait,
les résultats obtenus par dynamique moléculaire sont en parfait
accord avec ceux obtenus par la méthode de dynamique des contacts,
correspondant à des contacts parfaitement
rigides~\cite{Lemaitre08}.

La prise en compte de l'élasticité des grains conduit à tracer un
diagramme des régimes en fonction de $I/\sqrt{\kappa}$ et
$\Phi$~\cite{Campbell02}, mettant en évidence un régime
quasi-statique avec des effets de l'élasticité, un régime purement
inertiel sans effet de l'élasticité, et un régime intermédiaire
élastique-inertiel, qui ne s'avère accessible que pour des grains
très déformables ($\kappa < 100$).

\subsubsection{Influence du coefficient de frottement}

\begin{figure}[!htbp]
\begin{center}
\includegraphics*[width=\textwidth]{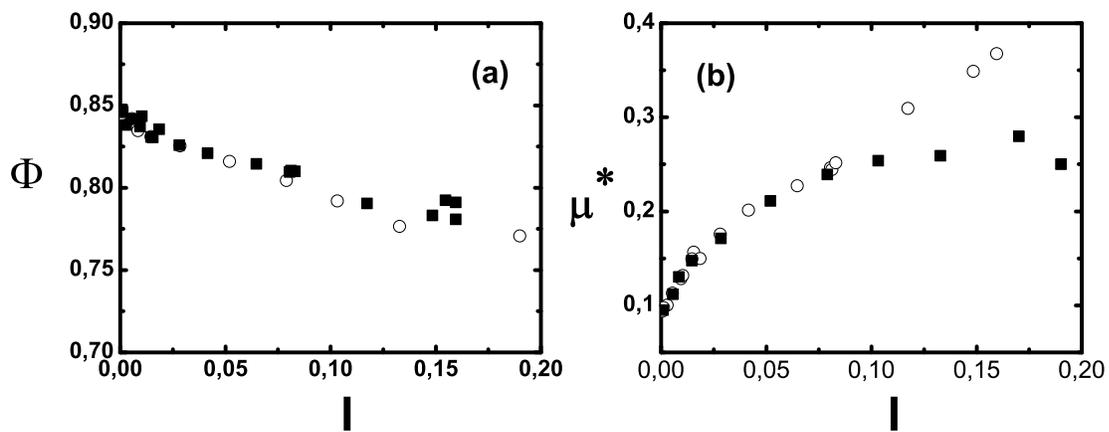}
\caption{\label{Fig.Ecoul2} \textit{(a) Loi de dilatance, (b) loi
de frottement ($\mu = 0$ - $e=0,1$ ($\circ$) - $e=0,9$ (carrés
noirs) - d'après~\cite{Dacruz05}.}}
\end{center}
\end{figure}

On a fait varier le coefficient de frottement $\mu$ entre $0$ et
$0,8$. On retrouve la même forme qualitative des lois de dilatance
et de frottement, avec des valeurs extrêmes $\Phi_{\text{max}}$ et $\mu
_{\min }^*$ dépendant sensiblement de $\mu$, conformément aux
variations des deux paramètres $\tg \phi$ et $\Phi_{c}$ de la loi
de comportement dans l'état critique, montrées au
§~\ref{sec:etatcritique}. La compacité reste une fonction linéaire
décroissante de $I$, avec $a \simeq 0,35$, et le frottement
effectif reste une fonction linéaire croissante de $I$, avec $b
\simeq 1$~\cite{Dacruz04a}. L'approximation linéaire est cependant
en défaut pour $I \le 0,01$ dans le cas des grains non frottants
(voir figure~\ref{Fig.Ecoul2}). La limite quasi-statique, dans de
tels matériaux, est approchée pour des valeurs beaucoup plus
faibles de $I$ que dans le cas de grains frottants, et
$\mu^*_{\text{min}}$ est lui-même beaucoup plus petit~\cite{Peyneau08}.
\subsubsection{Influence du coefficient de restitution}
On a fait varier $e$ entre $0,1$ et $0,9$~\cite{Dacruz04a}. Dans
le cas de grains frottants, le coefficient de restitution $e$ n'a
pas d'influence sur la loi de comportement
(figure~\ref{Fig.Ecoul1}). Dans le cas de grains non frottants, le
coefficient de restitution $e$ n'a pas d'influence sur la loi de
comportement pour $I<0,1$ (figure~\ref{Fig.Ecoul2}). Au-delà, le
comportement dépend de $e$ : lorsque $e \to 1$, la chute de la
compacité est moins marquée et le frottement effectif sature (voir
aussi~\cite{Campbell02}).
\subsubsection{Limite collisionnelle}
La situation $I >0,1 $, $\mu = 0$, $e = 0,9$ et $\eta=0$
correspond à la limite dense de la théorie cinétique, c'est à dire
des collisions binaires quasi-élastiques (la durée moyenne des
contacts devient égale au temps de collision~\cite{Dacruz04b}).
Dans cette limite dense, on déduit de la théorie
cinétique~\cite{Jenkins85b} la valeur du frottement effectif,
indépendant de $I$ mais dépendant de $e$ :
\begin{equation}
\mu^{*}(e) = \frac{1}{2} \left ( \frac{\pi + 8}{2\pi} \right
)^{\frac{1}{2}} \sqrt{1 - e^{2}}.
\end{equation}
Pour $e = 0,9$, on trouve $\mu^* = 0,29$ ce qui est en
assez bon accord avec la valeur mesurée. Cette évolution vers le
régime collisionnel explique ainsi la saturation du frottement
effectif pour $I> 0,1$ lorsque $e=0,9$ (figure~\ref{Fig.Ecoul2}b).
Ainsi, comme dans l'état critique, la loi de comportement des
écoulements denses ne dépend que de $I$ et de $\mu$ dans la limite
des contacts rigides ($\kappa$ assez grand). Il faut atteindre le
régime collisionnel pour que $e$ joue aussi un rôle.
\subsection{Influence du gradient de contrainte}
Il est indispensable de tester si la loi de comportement
identifiée en cisaillement homogène est valable dans d'autres
géométries de cisaillement où la distribution des contraintes est
hétérogène. On a ainsi considéré le cisaillement stationnaire et
uniforme dans la direction du cisaillement, dans des géométries
simples où l'on connaît la distribution des contraintes.

Sur plan incliné, une épaisseur $H$ de matériau s'écoule sur un
socle rugueux, incliné de $\theta$~\cite{Prochnow02, Dacruz04a}.
Dans ce cas, le coefficient de frottement effectif est constant
dans la couche en écoulement, égal à $\tg \theta$, de sorte que
la compacité et la nombre inertiel sont constants dans toute la
couche (sauf près du socle rugueux et de la surface libre). Au
coeur du matériau, on retrouve la loi de comportement mesurée en
cisaillement homogène. Le taux de cisaillement dépend du
coefficient de frottement~\cite{Azanza98, Silbert01}. Il est peu
sensible au coefficient de restitution, si ce dernier reste
inférieur à 0,7~\cite{Azanza98,Dippel99,Silbert01}. Cette
indépendance s'observe aussi vis à vis du coefficient de
restitution normal au socle sur toute la gamme des coefficients de
restitution~\cite{Campbell85a, Azanza98}.

En cisaillement annulaire~\cite{Koval09a}, le matériau est confiné
entre deux cylindres rugueux, intérieur de rayon $R_i$ et
extérieur de rayon $R_e$ assez grand ($\ge 2 R_i$) pour que la
paroi extérieure n'ait pas d'influence. Le cylindre intérieur
tourne à une vitesse de rotation $\Omega$. Le cylindre extérieur
exerce une pression radiale $P$. L'état de cisaillement est décrit
par $R_i$ et, à travers les valeurs imposées de $P$ et de
$\Omega$, par la vitesse tangentielle adimensionnée de la paroi
$V_{\theta} = \frac{\Omega R_i}{d} \sqrt{\frac{m}{P}}$, qui
généralise, à l'échelle du système global, la notion de nombre
inertiel introduite en cisaillement homogène. Une petite valeur de
$V_{\theta}$ correspond au régime quasi-statique, tandis qu'une
valeur élevée correspond au régime collisionnel. On utilise des
conditions aux limites périodiques en se restreignant à un secteur
angulaire $\Theta$ tel que $L=R_i \Theta$ soit supérieur à une
quarantaine de grains. Par ailleurs, les caractéristiques du
système simulé sont identiques à celles du système en cisaillement
homogène. Toutefois, si la pression radiale s'avère
approximativement constante, la contrainte de cisaillement décroît
comme l'inverse de la distance radiale $r$ au carré. Ceci, combiné
à des effets d'interfaces spécifiques~\cite{Koval08a}, conduit à
une localisation du cisaillement près du cylindre intérieur. La
mesure des relations entre les grandeurs locales $\Phi(r)$,
$\mu^*(r)$ et $I(r)$ montr que le comportement rhéologique ne
dépend pas de la distribution des contraintes dans le régime
inertiel. En revanche, lorsque $\mu^*$ devient inférieur à
$\mu^*_{\text{min}}$, les lois de dilatance et de
frottement~\eqref{eq:loisdilfrot} ne sont plus valables. On
observe un régime de fluage dont la compréhension reste un sujet
d'étude. Alors qu'en cisaillement homogène, $\mu^*_{\text{min}}$ est la
valeur maximale de $\mu^*$ supportée par le matériau granulaire
avant qu'il ne commence à s'écouler \emph{quasi-statiquement}, il
est capable de s'écouler sous ce seuil, lorsque la distribution de
contrainte est hétérogène.
\subsection{Influence de la cohésion}
\begin{figure}[!htbp]
\begin{center}
\includegraphics*[width=10cm]{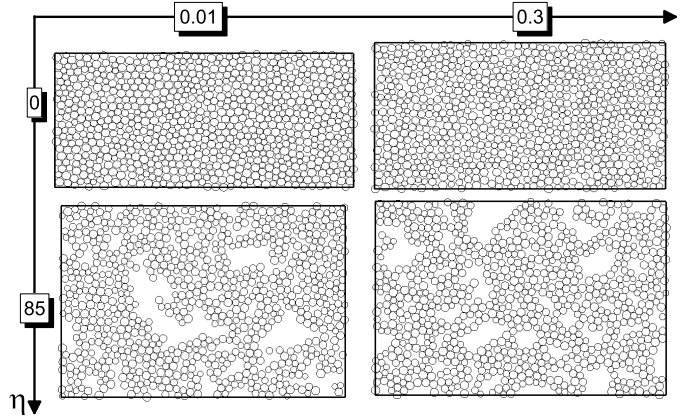}
\caption{\label{Fig.Ecoul3} \textit{Influence de $I$ et $\eta$ sur
la microstructure (d'après~\cite{Rognon08a}).}}
\end{center}
\end{figure}
\begin{figure}[!htbp]
\begin{center}
\includegraphics*[width=5cm]{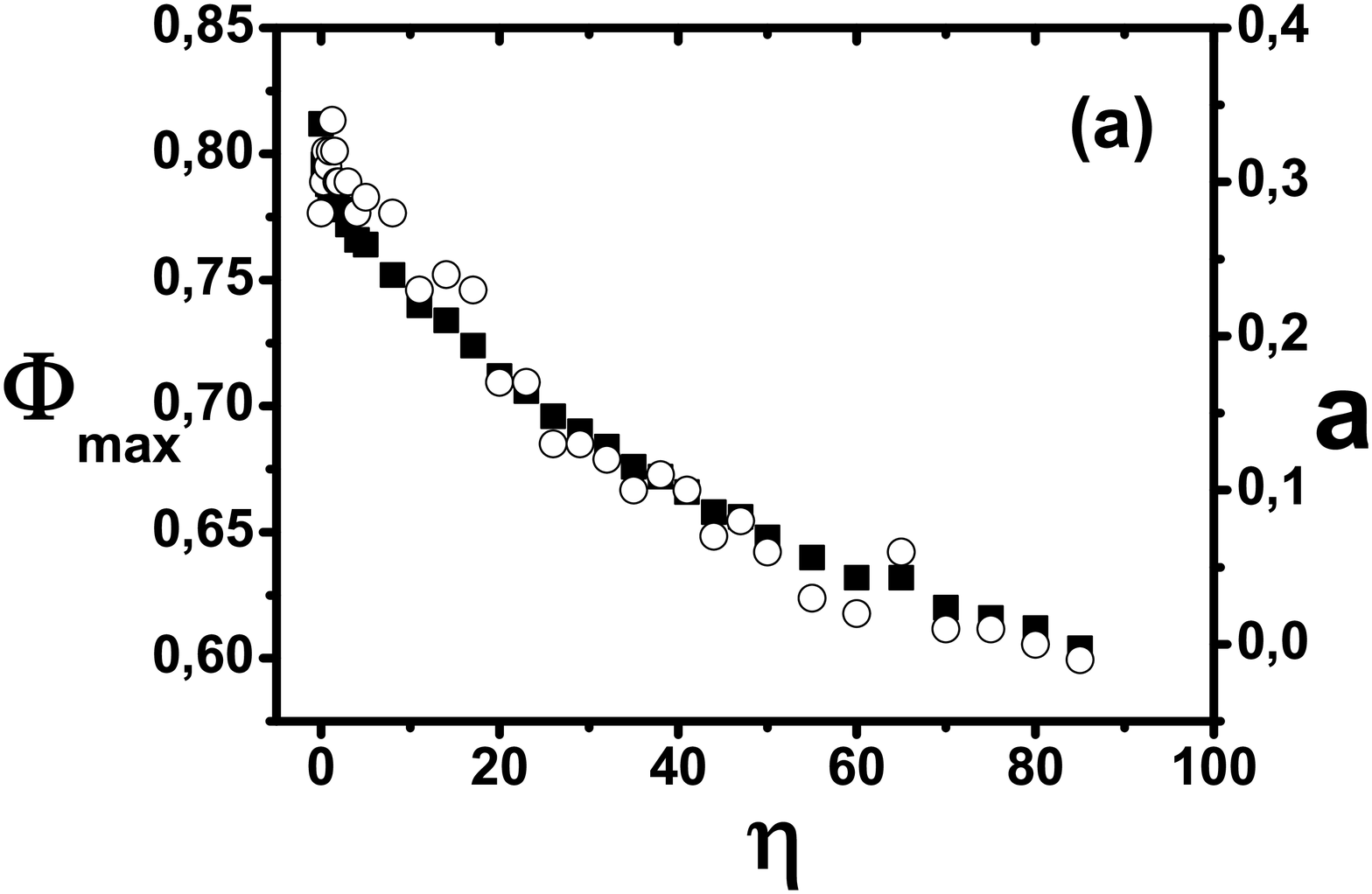}
\includegraphics*[width=5cm]{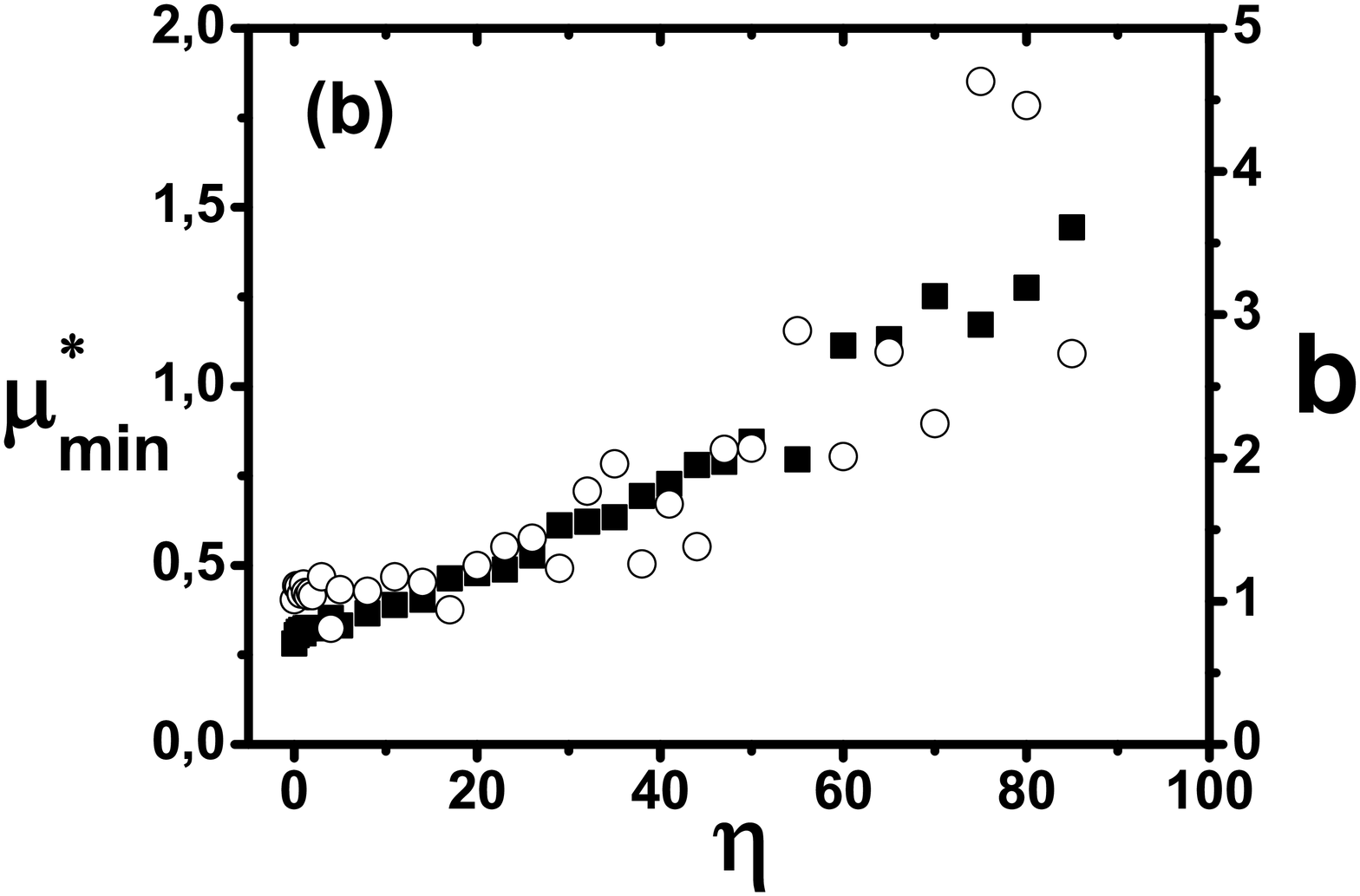}
\caption{\label{Fig.Ecoul4} \textit{(a) $\Phi_{\text{max}}(\eta)$ (carrés
noirs) et $a(\eta)$ $(\circ)$; (b) $\mu^{*}_{\text{min}}(\eta)$ (carrés
noirs) et $b(\eta)$ $(\circ)$ (d'après~\cite{Rognon06b}).}}
\end{center}
\end{figure}

Les résultats, issus de~\cite{Rognon06a,Rognon08a}, que nous rappelons ici font suite
au travail entamé par plusieurs
équipes~\cite{Mei00,Nase01,Radjai01c,Forsyth02,Weber04,Iordanoff05,Aarons06,Alexander06},
et visent à préciser quantitativement les lois rhéologiques en
présence de cohésion. En dehors de
l'introduction du nombre de cohésion $\eta$, les caractéristiques
du système simulé sont les mêmes que celles décrites au
§~\ref{sec:cishom}. On considère un modèle simple de force normale
de cohésion~\cite{Matuttis01,Radjai01c} qui prend en compte la
force attractive maximale $F_0$: $F^{a}(h) = -\sqrt{4 K_N F_0 h}$
(et $D_0=0$). En fait, la forme précise de la force attractive
$F^a(h)$ semble ne pas avoir d'influence sur le comportement
rhéologique, seul compte $F_0$\cite{Richefeu05,Rognon06a}. La
composante attractive de la force normale tend à rapprocher les
grains et la déformation moyenne $h^*$ dépend de l'intensité de la
cohésion $h^*(\eta) = \frac{1}{\kappa} \mathcal{H}(\eta)$ avec
$\mathcal{H}(\eta)= 1+2\eta+2\sqrt{\eta+\eta^2}$, d'ordre $1$ aux
faibles $\eta$ et d'ordre $4 \eta$ aux forts $\eta$. Le paramètre
$\kappa$ a été fixé à $10^{5}$, de sorte que l'on reste dans la
\emph{limite des contacts rigides} au moins pour les faibles
cohésion: $h^*(\eta) \le 10^{-4}$ pour $\eta \le 2,5$. L'influence
de la déformation des grains, pour les valeurs plus élevées de
$\eta$, est discutée dans \cite{Aarons06}. Lorsque l'intensité de
la cohésion n'est pas trop forte ($\eta < 100$), le matériau est
cisaillé de manière homogène (figure \ref{Fig.Ecoul3}). Les lois
de dilatance et de frottement~\eqref{eq:loisdilfrot} conservent la
même forme mais les paramètres $\Phi_{\text{max}}$, $\mu _{\min }^*$, $a$
et $b$ dépendent de $\eta$. La coh\'esion provoque une diminution
de la compacit\'e, mais une augmentation du nombre de
coordination, et une homog\'en\'eisation des directions de
contact. C'est le signe d'une organisation des grains en amas
compacts s\'epar\'es par du vide. La figure \ref{Fig.Ecoul4}a
trace les deux paramètres $\Phi_{\text{max}}$ et $a$ en fonction de
l'intensité de la cohésion $\eta$. Les deux fonctions ont une
forme similaire : une importante diminution dès les premiers
niveaux de cohésion ($\eta \le 2$) puis une diminution plus
faible. La diminution  de $a(\eta)$ jusqu'à $0$ correspond à des
systèmes dont la compacité ne dépend plus de l'état de
cisaillement $I$. Les deux fonctions $\mu^{*}_{\text{min}}(\eta)$ et
$b(\eta)$ ont la même forme (figure \ref{Fig.Ecoul4}b). Sous un
seuil de cohésion ($\eta \ge 10$), la cohésion n'affecte pas
$\mu^{*}_{\text{min}}$ ou $b$. Au-dessus du seuil, $\mu^{*}_{\text{min}}(\eta)$
et $b(\eta)$ augmentent fortement. On peut ainsi distinguer des
régime de faible ($\eta \le 10$) et forte ($\eta \ge 10$)
cohésion.

\begin{figure}[!htbp]
\begin{center}
\includegraphics[width= 7cm]{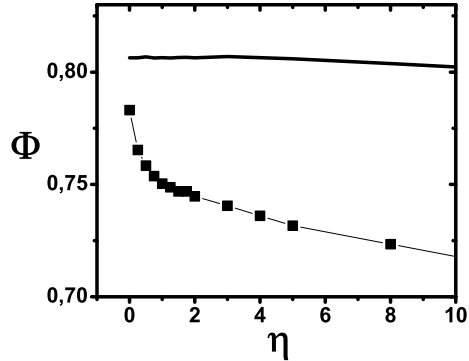}
\caption{\label{Fig.Ecoul5}\textit{$\Phi(\eta,I=0,2)$ pour des
grains frottants $\mu=0,4$ (carrés noirs) et non-frottants $\mu=0$
$(\circ)$ (d'après~\cite{Rognon08a}).}}
\end{center}
\end{figure}

%
%
%
La force de frottement entre les grains est décrite par un critère
de Coulomb qui considère uniquement la composante élastique de la
force normale. Écrit en fonction de la force normale totale, le
critère s'écrit $\vert\vert {\bf F}_T \vert\vert/F_N \le \mu
\mathcal{H}(\vert F_0/F_N \vert)$. Le coefficient de frottement
apparent en présence de cohésion est donc : $\mu \mathcal{H}(\vert
F_0/F_N \vert)$. Pour une faible cohésion, la force normale peut
être beaucoup plus grande que la résistance à la traction $F_N \gg
F_0$. Dans ce cas, la fonction $\mathcal{H}$ tend vers $1$ et le
coefficient de frottement apparent reste $\mu$. Pour une forte
cohésion, la force normale est beaucoup plus petite que la
résistance à la traction : $F_N \ll F_0$ et la fonction
$\mathcal{H}$ tend vers $4 \vert F_0/F_N \vert$. Par conséquent,
la cohésion augmente fortement le coefficient de frottement
apparent entre les grains, et comme ceci a été montré pour les
grains sans cohésion, ceci augmente la dilatance du matériau.
C'est ce que l'on mesure sur la figure~\ref{Fig.Ecoul5} qui
compare la compacité de grains frottants et non frottants
($\mu=0$).

\section{Conclusion}
Le choix de paramètres appropriés pour étudier des comportements mécaniques de matériaux granulaires
par simulation numérique discrète dépend bien sûr du problème particulier auquel on s'intéresse.
Il peut utilement s'appuyer sur l'analyse dimensionnelle et sur les connaissances que l'on a acquises, pour des matériaux
modèles simples dans différents régimes de comportement,
de l'influence des paramètres de contrôle sans dimension choisis au §~\ref{sec:anadim}~: les plus importants sont le nombre d'inertie
$I$ (défini en~\ref{eq:defI}), le niveau de raideur $\kappa$ (défini en~\ref{eq:defkappa}), et le cas échéant, le nombre de cohésion $\eta$
(défini en~\ref{eq:defI}).
Le choix de paramètres sans dimension retenu permet aussi commodément de traiter des matériaux solides comme en écoulement.
La loi constitutive en situation d'écoulement stationnaire prend la forme d'une << équation d'état >> qui donne l'ensemble
des caractéristiques mécaniques et de microstructure en fonction d'un seul paramètre ($I$) ou de deux ($I$ et $\eta$) pour les systèmes
cohésifs.
La table~\ref{tab:param} résume les résultats rapportés dans les §~\ref{sec:quasistatique} et~\ref{sec:ecoulement}. Aux trois nombres
$\kappa$, $I$ et $\eta$ (ou $P^*$) il faut
ajouter le coefficient de frottement intergranulaire $\mu$ (et éventuellement $\mu_R/d$, le frottement de roulement), dont la valeur influe
sur l'ensemble des comportements évoqués. On doit aussi garder à l'esprit le rôle crucial de la géométrie, celle des grains eux-mêmes, et celle
de l'état initial pour tous les comportements de type solide, sauf les états critiques.
Le choix des paramètres $I$ et $\kappa$ permet de
traiter de la limite quasi-statique et de celle des grains rigides, importantes en pratique. La valeur $I=10^{-3}$ indiquée dans la table pour
l'approche de la limite quasi-statique (comportement indépendant de $I$) est assez indicative, elle dépend bien sûr des systèmes étudiés
et des configurations auxquelles on s'intéresse. En général, la limite quasi-statique est d'autant plus facile à approcher que le système est
dense et bien coordonné, et demande davantage de précautions en présence de déformations de type II avec des réarrangements importants. De même,
la limite des grains rigides peut se situer un peu au-delà du seuil $\kappa\ge 1000$ indiqué.
L'effet du paramètre $\nu$ ou de
$K_T/K_N$ est limité au régime de déformations de type I, et faible si $K_T/K_N>1$.
\begin{table}
 \centering \small
\begin{tabular}{|c||c|c|c|c|c|c|}  \cline{1-7}
 & $I$ & $\kappa$ ({\scriptsize $>10^3$})&  $e$ ou $\zeta$& $\mu$& $\mu_R/d$& $\eta$\\
\hline \hline
{\scriptsize Assemblage} &O&N&O&O&O&O\\
\hline
{\scriptsize Solide, déf. type I, $I<10^{-3}$}&N&O&N&O&O&O \\
\hline
{\scriptsize Solide, déf. type II, $I<10^{-3}$}&N &N&N &O&O&O\\
\hline
{\scriptsize \'Etat critique, $I<10^{-3}$} &N &N&N&O&O&O \\
\hline
{\scriptsize \'Ecoulements denses, $10^{-3}<I<0.1$} &O&N&N&O&O&O\\
\hline
{\scriptsize R\'egime collisionnel, $I>0.1$} &O&N &O&O&O&O\\
\hline
\end{tabular}
\label{tab:param}
\normalsize
\caption{Influence des param\`etres sans
dimension sur les comportements m\'ecaniques dans différents régimes de sollicitation. (O=Oui,
N=Non)}
\end{table}
Dans cette limite, il peut être commode de modéliser les grains comme des solides indéformables, à la manière de la méthode
de dynamique des contacts (Chapitre 3). Cette approche peut donc s'appliquer à l'ensemble des cas recensés dans la table,
sauf le régime de déformation de type I. Celui-ci apparaît comme un domaine de comportement rigide
indéformable s'il est traité en dynamique des contacts.
Même si ce point mériterait des investigations complémentaires, on  s'attend à ce que l'instabilité des réseaux de contact qui borne l'intervalle
de comportement de type I coïncide en dynamique des contacts et dans une modélisation prenant en compte un niveau de raideur finie. L'approche
par des grains indéformables est alors légitime tant que l'on peut ignorer les déformations de l'ordre de quelques $\kappa^{-1}$ qui surviennent
dans le matériau réel.

Les processus d'assemblage figurent dans la table, mais il conviendrait bien sûr de
distinguer les différentes méthodes de préparation d'échantillons solides (chapitre 8). Leur
description précise demande d'introduire d'autres paramètres de contrôle.
Comme il s'agit souvent de procédés impliquant le passage d'un état
agité et lâche à un assemblage solide en équilibre, les paramètres dynamiques $I$ et $e$
sont susceptibles d'influencer le résultat (comme dans le cas de la pluviation contrôlée~\cite{Emam06}).
La compression de matériaux solides dans un trajet de chargement conservant le rapport des contraintes, comme
évoqué au §~\ref{sec:comp}, est à classifier avec le comportement solide en déformation de type I en l'absence de
cohésion, et comme déformation de type II dans le cas cohésif (on évaluera alors, si $\eta>1$, les effets dynamiques
avec $I_a$). En présence de cohésion, tous les comportements sont sensibles à la valeur de $\eta$, sauf bien sûr
dans la limite $\eta\to 0$ (pour la consolidation $\eta < 5. 10^{-2}$ semble suffisant pour négliger la cohésion).

Au-delà de son utilité pratique pour le choix de paramètres appropriés dans une application donnée de la simulation discrète,
la classification des paramètres de contrôle et de leur domaine d'influence sur les comportements est également
précieuse pour la compréhension des origines microscopiques des lois constitutives macroscopiques.

\bibliographystyle{unsrt}
\bibliography{chapter9_biblio}

\end{document}